\documentclass{article}  

\usepackage{graphicx}

\usepackage{geometry}
\usepackage{txfonts}
\usepackage{amssymb}
\usepackage{amstext}
\usepackage{natbib}

\usepackage{hyperref}
\newcommand{\jgr}{J. Geophys. Res. }
\newcommand{\grl}{Geophys. Res. Lett. }
\newcommand{\icarus}{Icarus }
\newcommand{\aap}{Astron. Astrophys. }
\newcommand{\apj}{Astrophys. J. }
\newcommand{\apjl}{Astrophys. J. Lett. }

\newcommand{\aj}{Astron. J. }
\newcommand{\planss}{Planet. Space Sci. }
\newcommand{\araa}{Ann. Rev. Astron. Astrophys. }
\newcommand{\ssr}{Space Sci. Rev. }
\newcommand{\mnras}{Mon. Not. R. Astron. Soc. }
\newcommand{\nat}{Nature }
\newcommand{\science}{Science }

\newcommand{\jchemphys}{J. Chem. Phys. }
\newcommand{\jqsrt}{J. Quant. Spectrosc. Radiat. Transfer }

\newcommand{\natastron}{Nat. Astron. }

\newcommand{\psj}{Planet. Sci. J. }
\newcommand{\areps}{Annu. Rev. Earth Planet. Sci. }

\newcommand{\natcom}{Nat. Commun.} 
\newcommand{\ptrslsA}{Philos. Trans. R. Soc. A} 

\newcommand{\kms}{\,km$\cdot$s$^{-1}$}
\newcommand{\ms}{\,m$\cdot$s$^{-1}$}

\newcommand{\ten}[1]{$10^{#1}$}

\newcommand{\unitgamma}[0]{cm$^{-1}\cdot$atm$^{-1}$}

\newcommand{\Fluxunit}{\,cm$^{-2}\cdot$s$^{-1}$}
\newcommand{\Kunit}{\,cm$^{2}\cdot$s$^{-1}$}
\newcommand{\Coldens}{\,cm$^{-2}$}
\newcommand{\dix}[1]{$\times10^{#1}$}
\newcommand{\fig}[1]{Fig.~\ref{#1}}
\newcommand{\figs}[2]{Figs.~\ref{#1} and \ref{#2}}

\newcommand{\tab}[1]{Table~\ref{#1}}

\newcommand{\degre}{\ensuremath{^\circ}}

\begin{document} 


\Large
\begin{center}\textbf{Detection of stratospheric HCN and tropospheric CO in Uranus and the implication for their sources}\end{center}\normalsize

\large\noindent T. Cavali\'e$^{1}$, R. Moreno$^{2}$, C. Lefour$^{1}$, S. H. Luszcz-Cook$^{3,4}$, T. Fouchet$^{2}$, E. Lellouch$^{2}$, O. Carri\'on-Gonz\'alez$^{5,2}$, B. Benmahi$^{6,7}$, S. Guerlet$^{8,2}$, G. Milcareck$^{9}$,  N. Cl\'ement$^{10}$, M. Dobrijevic$^{1}$, I. de Pater$^{11}$, A. Moullet$^{12}$, J. I. Moses$^{13}$, J. Leconte$^{1}$, A. Spiga$^{8}$, M. T. Roman$^{14,15}$, L. N. Fletcher$^{15}$\normalsize\\
\vspace{0.2cm}

\noindent$^1$Laboratoire d'Astrophysique de Bordeaux, Univ. Bordeaux, CNRS, B18N, all\'ee Geoffroy Saint-Hilaire, 33615 Pessac, France (ORCID: 0000-0002-0649-1192, email: thibault.cavalie@u-bordeaux.fr)\\ 
$^2$LIRA -- Laboratoire d'Instrumentation et de Recherche en Astrophysique, Observatoire de Paris, Section de Meudon, 5, place Jules Janssen -- 92195 MEUDON Cedex\\
$^3$Liberal Studies, New York University, New York, NY 10003, USA\\
$^4$Department of Astrophysics, American Museum of Natural History, New York, NY 10024, USA\\
$^5$Max-Planck-Institut f\"ur Astronomie, K\"onigstuhl 17, 69117 Heidelberg, Germany\\
$^6$Aix-Marseille Universit\'e, CNRS, CNES, Institut Origines, LAM, Marseille, France\\
$^7$Atmospheric and Planetary Physics, STAR Institute, University of Li\`ege, Belgium\\
$^8$Laboratoire de M\'et\'eorologie Dynamique/Institut Pierre-Simon Laplace (LMD/IPSL), Sorbonne Universit\'e, CNRS, \'Ecole Polytechnique, Institut Polytechnique de Paris, \'Ecole Normale Sup\'erieure (ENS), PSL Research University, Paris, France\\
$^9$Instituto de Astrof\'isica de Andaluc\'ia (IAA), CSIC, Granada, Spain\\
$^10$Institut Pierre-Simon Laplace, Sorbonne Universit\'e/CNRS, Paris, France\\
$^11$Department of Astronomy, Department of Earth and Planetary Science, University of California, Berkeley, CA 94720, USA\\
$^12$National Radio Astronomy Observatory, Charlottesville, VA 22903, USA\\
$^13$Space Science Institute, 4765 Walnut Street, STE B, Boulder, CO 80301, USA\\
$^14$Facultad de Ingeniera y Ciencias, Universidad Adolfo Ib\'a\~nez, Av. Diagonal las Torres 2640, Pe\~nalol\'en, Santiago 7941169, Chile\\
$^15$School of Physics and Astronomy, University of Leicester, Leicester LE1 7RH, UK\\

\vspace{0.2cm}
\noindent\textbf{Received:} 29 April 2026\\
\noindent\textbf{Accepted:} 22 May 2026\\
\vspace{0.2cm}

\noindent\textbf{DOI:} 10.1051/0004-6361/202660695 \\
\vspace{0.5cm}

\section*{Abstract}
   Uranus belongs to the category of ice giants that are common in our Galaxy. However, Uranus is one of the least explored and understood planets in our Solar System. This work aims to constrain the deep oxygen abundance of Uranus to better understand its formation. Another goal concerns the origin of exogeneous species, such as CO and HCN, found in the upper stratospheres of giant planets. We used spectral mapping observations of the CO (J=3-2) and HCN (J=4-3) rotational submillimeter lines obtained with ALMA in 2022 and 2024. We combined them with radiative transfer and thermochemical modeling to determine the tropospheric abundance of CO and the deep O/H ratio of Uranus. We used radiative transfer simulations with physical models of various sources of external CO and HCN to constrain the vertical and meridional distributions of these species and narrow down the nature of their external sources. We also applied a wind retrieval algorithm to search for zonal winds in the stratosphere of Uranus at the levels probed by the CO and HCN lines. We unambiguously detect tropospheric CO for the first time with a mole fraction of 5.8$\pm$0.3\,ppb and stratospheric HCN with a mole faction of (1.8$\pm$0.2)\dix{-11} restricted to pressures lower than 0.2\,mbar. Thermochemical calculations suggest that the deep interior of Uranus is enriched in oxygen with respect to the protoplanetary nebula by at least a factor of 52$^{+30}_{-20}$. We also find that the stratospheric CO is rather uniform over the observed latitudes and that the CO lines are best fit by an old comet impact model, in which a large comet hit the planet several centuries ago. CO therefore has a dual origin in Uranus. Finally, we do not detect stratospheric winds from these data, but the CO data indicate that zonal winds in the 10$^\circ$S-10$^\circ$N latitudinal range are likely retrograde at submillibar pressures.

\section{Introduction}
Over the last three decades, exoplanet research has revealed that gas giants -- especially Neptune-class planets -- are abundant throughout our galaxy \citep{Zhu2021}. Consequently, the giant planets within our own Solar System, particularly the ice giants, provide essential benchmarks for studying how these distant worlds form and evolve. 

To unveil how giant planets were formed and how they work, we have to measure their composition, structure, dynamics, seasonal evolution, etc. Contrary to Jupiter and Saturn, the exploration of Uranus and Neptune remains limited. The two planets have only been visited by the Voyager 2 spacecraft in 1986 and 1989, respectively. Despite regular observations with Earth-based and space telescopes, many unsolved questions remain with regard to their composition and dynamics \citep{Atreya2020,Fletcher2020a,Fletcher2020b,Guillot2023,Helled2020b,Moses2018}. In this paper, we explore the composition of the atmosphere of Uranus with the aim of constraining the vertical and meridional distributions of CO and HCN and their internal and external sources. 

While abundant CO is present in Neptune's troposphere ($\sim$0.2\,ppm, \citealt{Hesman2007,Luszcz-Cook2013,Teanby2019,Moreno2017}), CO has surprisingly not been detected in the troposphere of Uranus yet, and the current upper limit set by \citet{Teanby2013} is 3 orders of magnitude lower than in Neptune, potentially implying vastly different O/H ratios. This difference is puzzling in many aspects. It raises the following fundamental questions: (i) whether or not Uranus and Neptune formed from different types of building blocks. Several scenarios can explain different elemental compositions for Uranus and Neptune, depending, for example, on their precise formation location with respect to the various volatile snowlines \citep{Mousis2020}; (ii) If Uranus and Neptune share similar formation material and processes, whether or not their interiors evolved differently. The negligible intrinsic luminosity of Uranus compared to the much higher one of Neptune \citep{Guillot2005,Irwin2025} may result from different internal structures. Compositional gradients, possibly caused by early giant impacts \citep{Reinhardt2020}, may inhibit convection deep in the interior of Uranus \citep{Vazan2020,Helled2020a}, meaning that any deep CO is not efficiently mixed upward to the levels accessible to remote sensing.
 
Higher up in the stratosphere, the detection of oxygen species such as H$_2$O, CO, and CO$_2$ in all four giant planets and Titan is linked to the infall of oxygen-bearing molecules from their environment \citep{Feuchtgruber1997,Coustenis1998,Cavalie2010,Cavalie2014,Burgdorf2006,Lellouch2005}. In addition, other species such as HCN and CS have also been detected and proven to have an external origin in Jupiter \citep{Lellouch1995,Marten1995,Moreno2003,Cavalie2023b} and Neptune \citep{Lellouch2005,Luszcz-Cook2013,Moreno2017}. These species can be delivered to the giant planets by: (i) interplanetary dust particles (IDPs) \citep{Landgraf2002,Poppe2016,Moses2017,Moses2000b}, (ii) material originating from icy rings and/or satellites being transported inward in the system to rain down on the planet  \citep{Strobel1979,Prange2006,Hartogh2011a,Waite2018,Perry2018,Cavalie2019}, (iii) large comet impacts \citep{Lellouch1995,Lellouch2005,Cavalie2010,Luszcz-Cook2013}, or (iv) a combination of those. For CO and HCN, an internal source can also contribute to stratospheric abundances, because CO does not condense at the tropopause and can be transported up to the stratosphere, and because HCN can be produced from the photochemistry of internally sourced N$_2$.

The three possible infall pathways lead to specific signatures in terms of chemical composition and spatio-temporal and vertical distributions of the various infall species. For instance, comets are sporadic and spatially localized \citep{Lellouch1995}, IDPs are steady, latitudinally and longitudinally uniform \citep{Moses2017}, and ring and satellite sources are localized \citep{Waite2018,Cavalie2019}. These differences have been used to constrain the origin of external sources on Jupiter \citep{Lellouch2002,Lellouch2006,Cavalie2013}, Saturn \citep{Cavalie2009,Cavalie2010,Cavalie2019}, and Neptune \citep{Lellouch2005,Luszcz-Cook2013,Moreno2017}. At Uranus, H$_2$O probably comes from IDPs \citep{Moses2017,Teanby2022}. The dominant source of CO in Jupiter, Saturn, and Neptune seems to be comet impacts \citep{Moreno2003,Cavalie2010,Lellouch2005}. On Uranus, \citet{Encrenaz2004} favored an external source from their first detection of the CO in the stratosphere around 4.7\,$\mu$m. \citet{Teanby2013} did not detect CO with Herschel/SPIRE, but the upper limit they derived for an internal component strongly suggested that the majority of the CO observed by \citet{Encrenaz2004} had an external origin. Using Herschel/HIFI, \citet{Cavalie2014} confirmed the external origin of stratospheric CO, but its nature has remained unknown ever since, even if photochemical modeling indicate that comets presumably provided an important source of the oxygen on Uranus \citep{Moses2017}. Three dimensional maps of infall species such as CO have the potential to distinguish between the possible delivery mechanisms.

The stratospheric signatures of molecules such as, for example, H$_2$O, CO, HCN, and CS, when observed with sufficient spatial and spectral resolution can be further used to constrain the circulation in these atmospheric layers. Wind speeds can be determined from the measurement of the Doppler shifts induced by the winds on the lines, even though these altitude layers are devoid of the usual wind tracers, i.e., the clouds. This technique has now been successfully applied to the giant planets and Titan, with the exception of Uranus \citep{Cavalie2021,Cavalie2026,Benmahi2021,Benmahi2022,Benmahi2025,Carrion-Gonzalez2023,Moreno2005,Lellouch2019}. In Uranus, cloud-top wind measurements \citep{Hammel2005,Sromovsky2015,Sromovsky2024,Hueso2020} show a broad retrograde flow within $\pm$30$^\circ$ of latitude around the equator, with a peak velocity of $\sim$-50\ms~at the equator. At higher southern and northern latitudes, the flows are prograde with top speeds of $\sim$$+$250\ms~at 60$^\circ$ latitude. Wind speeds in the stratosphere can, in principle, be inferred from the thermal wind equation \citep{Flasar1987,Fletcher2020a}, although this remains uncertain given the absence of discrete stratospheric tracers. Measuring directly the wind speeds in the stratosphere is therefore key to understanding how momentum is transported from the weather layer to upper atmospheric layers, and constraining general circulation and equatorial oscillations. This is particularly true for Uranus and Neptune considering their long and extreme seasons \citep{Milcareck2024}.

With this paper, we aim at constraining the internal and external sources of CO on Uranus from spatially resolved spectral mapping with Atacama Large Millimeter/Submillimeter Array (ALMA). We also present the first detection of HCN in the stratosphere of the planet. In addition, we use the signatures of CO and HCN in the stratosphere to tentatively constrain wind speeds at the probed altitudes. We present the observations and models in Sections \ref{sec:Observations} and \ref{sec:Models}, respectively. We give our results regarding the vertical and meridional distribution of CO in the troposphere and in the stratosphere (respectively) of Uranus in Sections \ref{sec:Internal} and \ref{sec:Meridional} (respectively). We discuss the origin of CO in the atmosphere of the planet in Section \ref{sec:Combined}. We show the first detection of HCN, constrain its stratospheric abundance, and discuss its origin in Section \ref{sec:HCN}. We derive constraints on stratospheric winds from CO and HCN observations in Section \ref{sec:Winds}. Finally, we give our conclusions in Section \ref{sec:Conclusion}.

\section{Observations}\label{sec:Observations}
Uranus was observed with ALMA on August 19 and October 18, 2022, and on July 30, 2024, as part of projects 2021.1.01034.S and 2022.1.00558.S (PI: S. Luszcz-Cook). The spectral setup enabled the simultaneous observation of the CO (J=3-2) transition at 345.796\,GHz and the HCN (J=4-3) transition at 354.505\,GHz over bandwiths of 2\,GHz and with a resolution of 1.1 \,MHz. The integration time on Uranus was around 40 minutes on each date. All geometrical details, as retrieved from the JPL Horizons database\footnote{\url{https://ssd.jpl.nasa.gov/horizons/}} for the relevant dates, are given in \tab{tab:geometry}. Details of each ALMA configuration are given in \tab{tab:ALMA-obs}. The planet was observed with the C-3 subcompact and moderately extended C-5 configurations, providing a spatial resolution comprised between 0.2'' and 0.7'' onto a $\sim$3.6'' disk. The planet spatial resolution (planet size / beam size) obtained on the three dates is then 17, 6, and 18, respectively. For bandpass, flux, and phase, the following nearby quasars were observed: J0238+1636 and J0252+1718 on August 19, 2022, J0309+1029 on October 18, 2022, and J0325+2224 on July 30, 2024. Atmospheric calibration was achieved with a water vapor radiometer.

After retrieving the pipeline calibrated data from the ALMA archive, we proceeded with continuum and line imaging of the individual datasets using the Common Astronomy Software Applications (CASA) version 6.4.1-12 for the October 18, 2022, data and CASA 6.5.4-9 for the other data, as recommended by the observatory. The continuum images were obtained by averaging all the continuum spectral channels and cleaning the data using the Clark deconvolution algorithm with Briggs weighting and a robust factor of 0.5 over several ten thousands of iterations combined with an elliptical mask encompassing the planet. The final images were corrected for the primary beam. The spectral line cubes were obtained after subtracting the continuum from the visibilities (uv data) and cleaning all spectral channels following the same method as for the continuum images. The continuum subtraction is a required step when imaging a faint line over a strong continuum source. In a final post-processing step, we added back the measured continuum to the spectral cubes to have line$+$continuum spectral images, directly comparable with radiative transfer outputs. 

  \subsection{CO}
\begin{figure*}[!h]
    \centering
    \includegraphics[width=8cm]{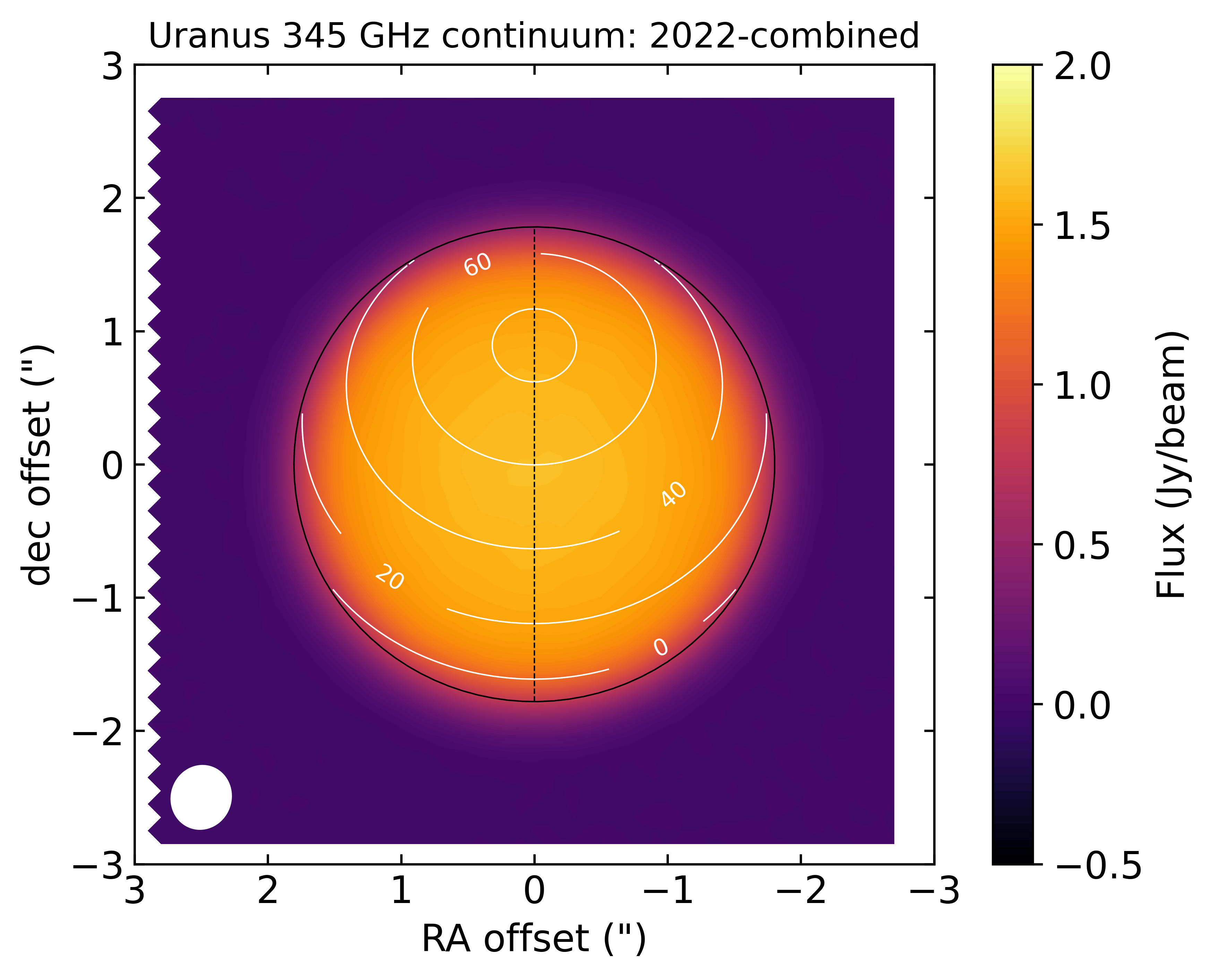}
    \includegraphics[width=8cm]{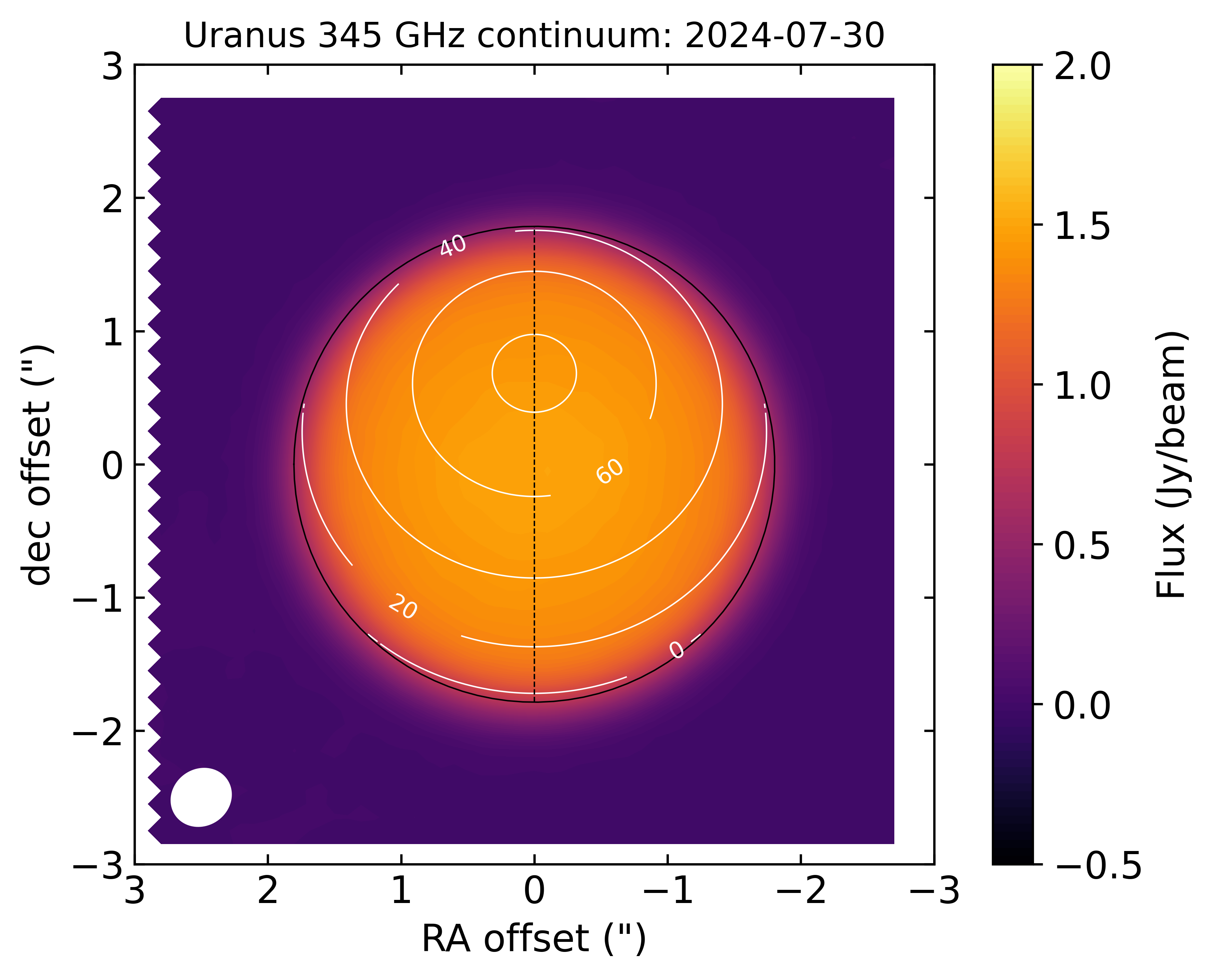}
    \caption{Uranus continuum images at 345\,GHz. The image on the left was obtained from the combined-2022 observations (August 19 and October 18). The one on the right corresponds to the data from July 30, 2024, alone. The planet 1-bar level is delimited by the black ellipse, latitudes are indicated with white isocontours, the central meridian is depicted by the dashed black line, and the beam is represented by the filled white ellipse.}
    \label{fig:continuum}
  \end{figure*}

  To improve the signal-to-noise ratio (S/N) of the data, we combined the visibilities of the different observations into two different datasets. We first coadded all the 2022 data together in a ``combined-2022'' dataset. The two observations were performed at close enough times that the sub-Earth point remained the same within 1\degre~of latitude. The combined-2022 data therefore has the highest S/N at the limb of Uranus, maximizing our sensitivity to the meridional distributions of CO. We also combined all the 2022 and the 2024 data together in a ``2022$+$2024'' dataset. Here, the goal is the optimization of the sensitivity to any disk center absorption feature that would be caused by tropospheric CO. To construct these two combined datasets, we exported the visibilities from the different datasets and rescaled the respective visibilities in the uv-plane to a common Uranus disk size of 3.61''. We then combined the rescaled uv tables to produce our combined-2022 and 2022$+$2024 datasets. To further increase the S/N in the cubes, we applied a uv-taper such that the final synthesized beam would have an almost circular diameter of $\sim$0.45'' (see \tab{tab:ALMA-obs}) and proceeded with the imaging using the Gildas\footnote{\url{https://www.iram.fr/IRAMFR/GILDAS/} and \url{https://ascl.net/1305.010}} software \citep{Pety2005}. Phase self-calibration was also performed with a continuum model derived from our radiative transfer model (see Section \ref{sec:Models}). The continuum flux was rescaled to the model, which has an absolute accuracy of 5\%. Finally, the images were derotated by the North Pole (NP) angle in order to have the North Pole pointing up. The final continuum images are displayed in \fig{fig:continuum}.
  
\begin{figure}[!h]
    \centering
    \includegraphics[width=12cm]{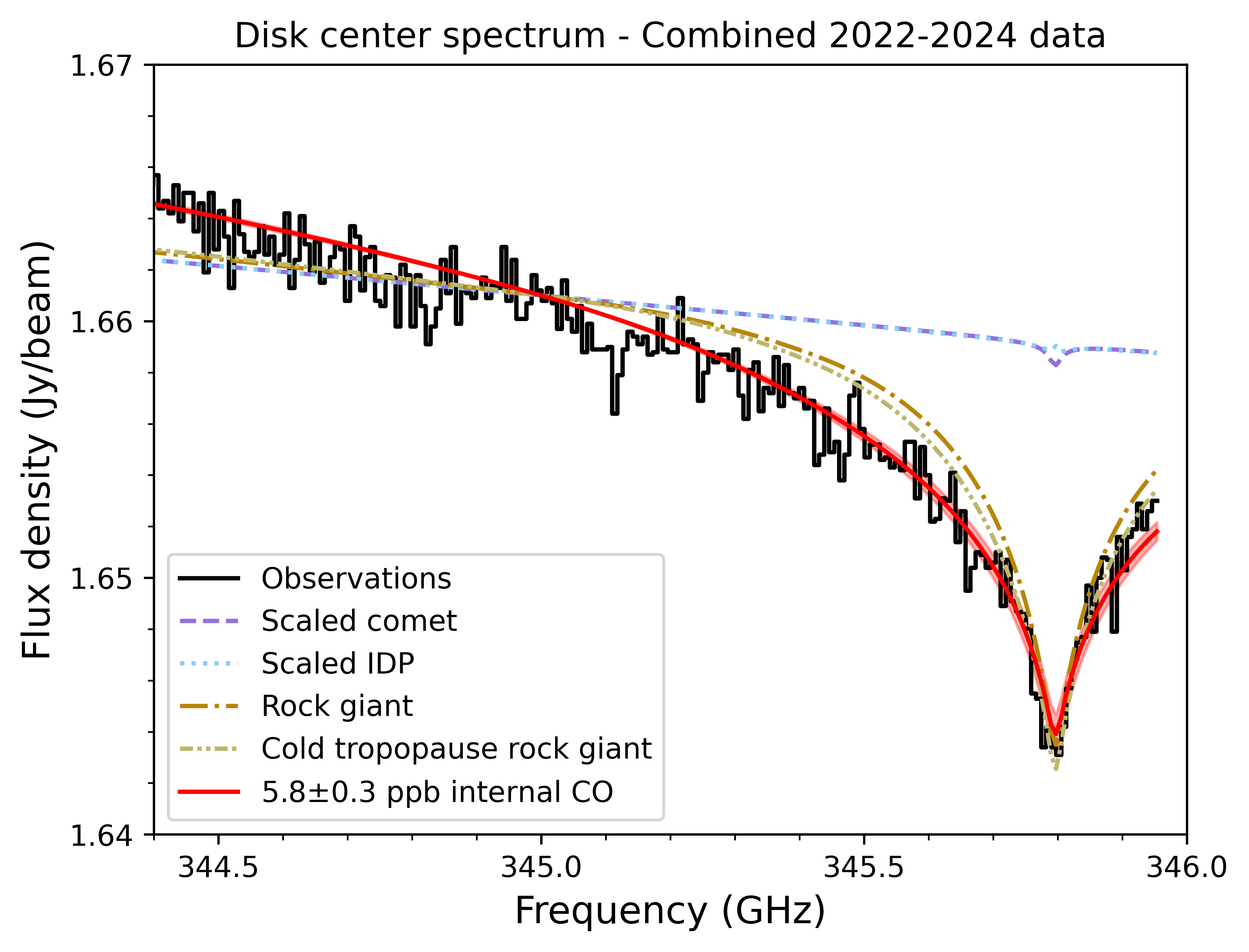}
    \caption{Uranus disk-center spectrum at 345\,GHz, as extracted from the 2022$+$2024 spectral cube. The spectral resolution is 8.8\,MHz. The ALMA bandwidth does not cover the full line. The tropospheric absorption can be reproduced with a deep CO mole fraction of (5.8$\pm$0.3)\dix{-9}. The best-fit models for external sources discussed in Section \ref{sec:Meridional} cannot reproduce such a broad absorption nor can a ``rock giant'' model or a "rock giant" model with a 5\,K colder tropopause region.}
    \label{fig:CO-disk-center}
  \end{figure}

\begin{figure*}[!h]
    \centering
    \includegraphics[width=8cm]{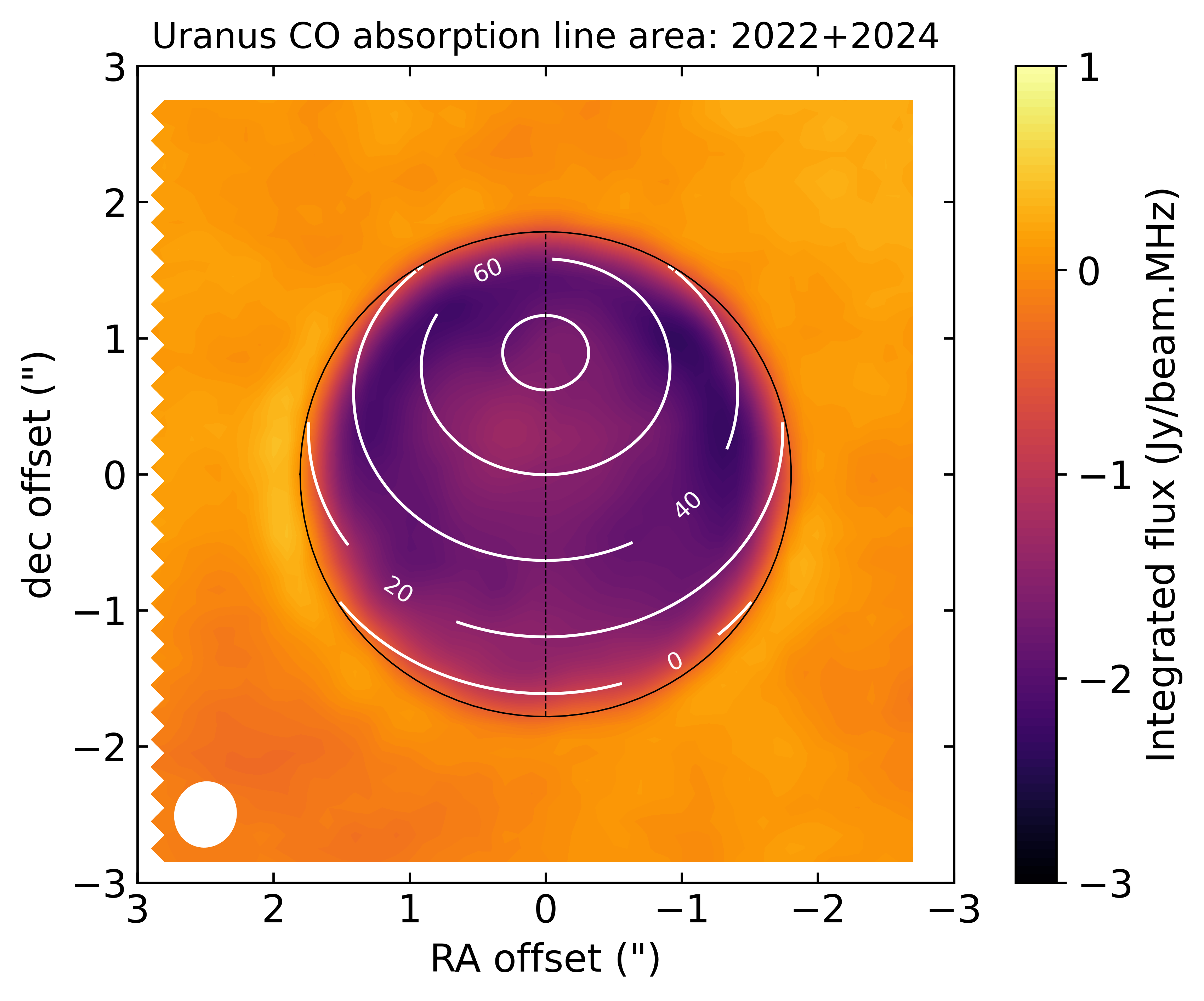}
    \includegraphics[width=8cm]{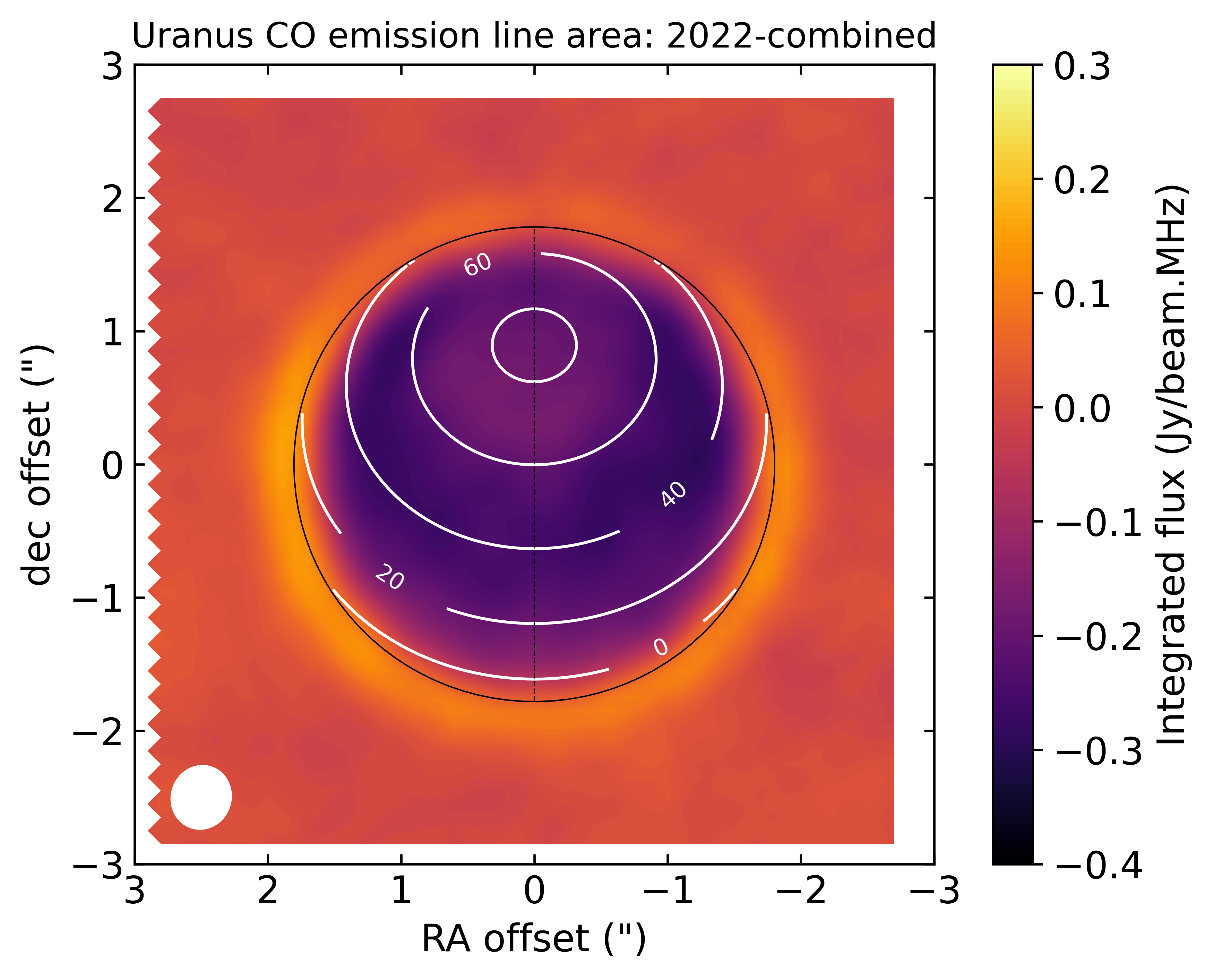}
    \includegraphics[width=8cm]{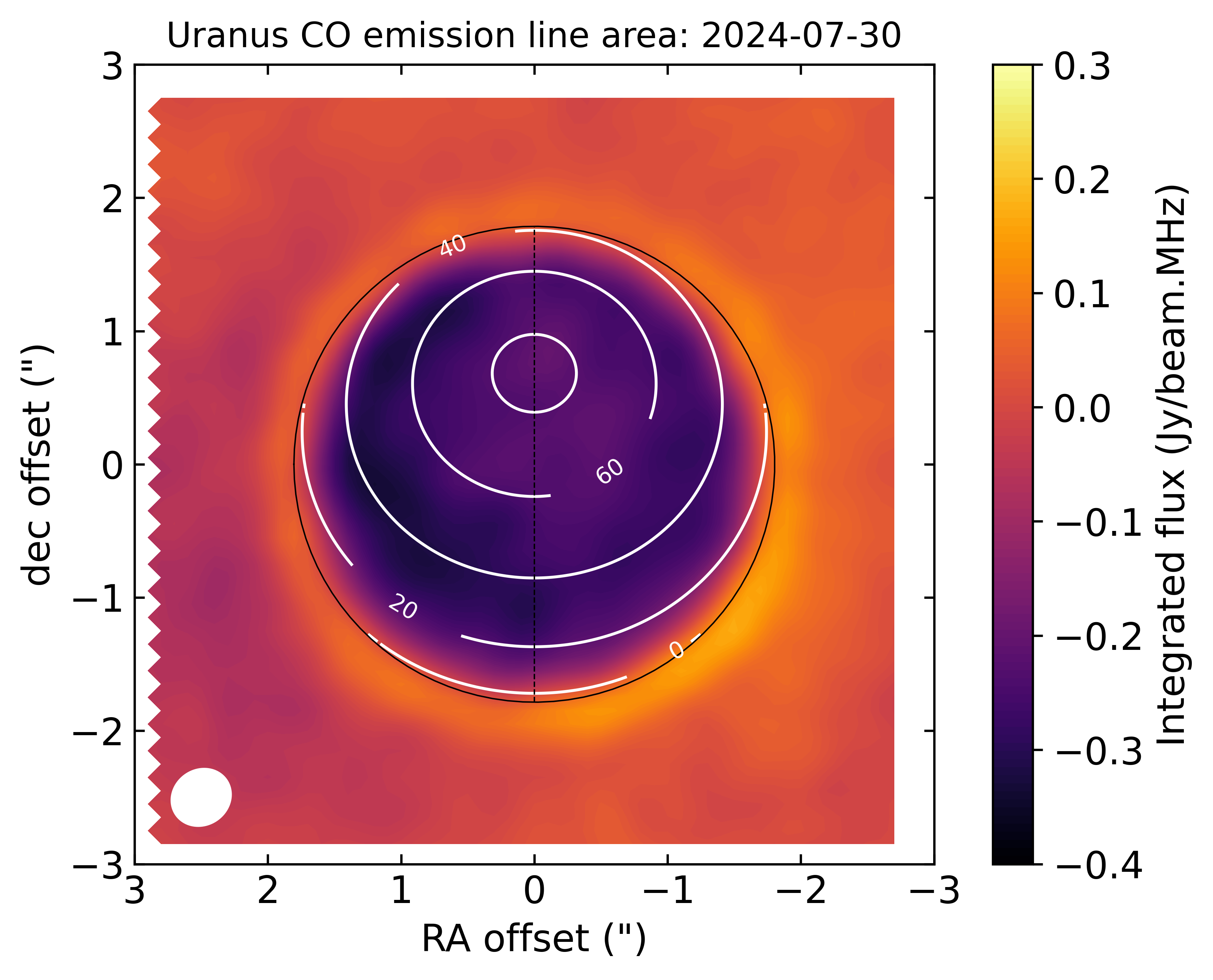}
    \caption{Uranus CO maps. (Top left) Uranus CO (J=3-2) broadband (from -140\kms~to 200\kms) area map from the 2022$+$2024 observations. This illustrates the disk center absorption discussed in Section \ref{sec:Internal}. (Top right) Uranus CO (J=3-2) emission line area image, obtained from the combined-2022 observations (August 19 and October 18). (Bottom) Uranus CO (J=3-2) emission line area image, obtained from the July 30, 2024, data alone. The latter two illustrate the stratospheric emission discussed in Section \ref{sec:Meridional} detected at the limb by limiting the flux integration from -10\kms~to +10\kms. The layout is the same as in \fig{fig:continuum}. Note that the latitudes of 2022 are only given for illustration for the 2022$+$2024 cube on the left, because this cube mixes observations with sub-Earth latitudes and North Pole angles that both differ by 7-8$^\circ$ between 2022 and 2024.}
    \label{fig:linearea}
  \end{figure*}

  The broad absorption caused by tropospheric CO is unambiguously detected for the first time in those data. This is demonstrated both with the disk center spectrum shown in \fig{fig:CO-disk-center} and with the line integrated map, in which the CO line flux density is integrated over velocity channels around line center (from $-140$\kms~to $+200$\kms, equivalent to 345.57-345.96\,GHz; see \fig{fig:linearea} left). The spectrum of \fig{fig:CO-disk-center}, in which we applied spectral averaging by a factor of 8 (i.e., spectral resolution of 8.8\,MHz) to increase the S/N, is reliable over most of the observed frequency range, with the exception of the lower end. Between 344.2 and 344.4\,GHz, a broad feature is consistently observed in all the pixels of the map, inside and outside of the planet. This feature is caused by imperfect terrestrial atmospheric calibration in this spectral region. We consequently ignore this spectral region in what follows.
  
  When integrating the line flux density in the combined-2022 and in the 2024 spectral images over velocity channels ranging from -10\kms~to +10\kms, we obtain the line area maps shown in \fig{fig:linearea} (center and right). The line emission is clearly concentrated at the limb. This is caused by the combination of an increased CO opacity due to longer path at the limb and a subdued continuum emission there. We can thus constrain the stratospheric profile of CO by analyzing the spectra at these locations. We therefore extracted spectra along the limb, defined as the 1-bar level, from each spectral cube and oversampled the beam by a factor of 4. This resulted in the production of 106 and 112 spectra, respectively from the combined-2022 and 2024 cubes, and enabled the oversampling of the latitudinal resolution of the observations. Having detections of the stratospheric emission line only at the limb results in one of the main limitations of this work for what concerns the characterization of the external source: only a limited range of latitudes is probed because of the high sub-observer latitude for each observation (see \tab{tab:geometry}). The ranges of latitudes probed in the two cubes are 17$^\circ$S-42$^\circ$N, and 10$^\circ$S-34$^\circ$N, respectively. We note that all latitudes given in the paper are planetocentric.

  \subsection{HCN \label{sec:HCN-obs}}
\begin{figure}[!h]
    \centering
    \includegraphics[width=9cm]{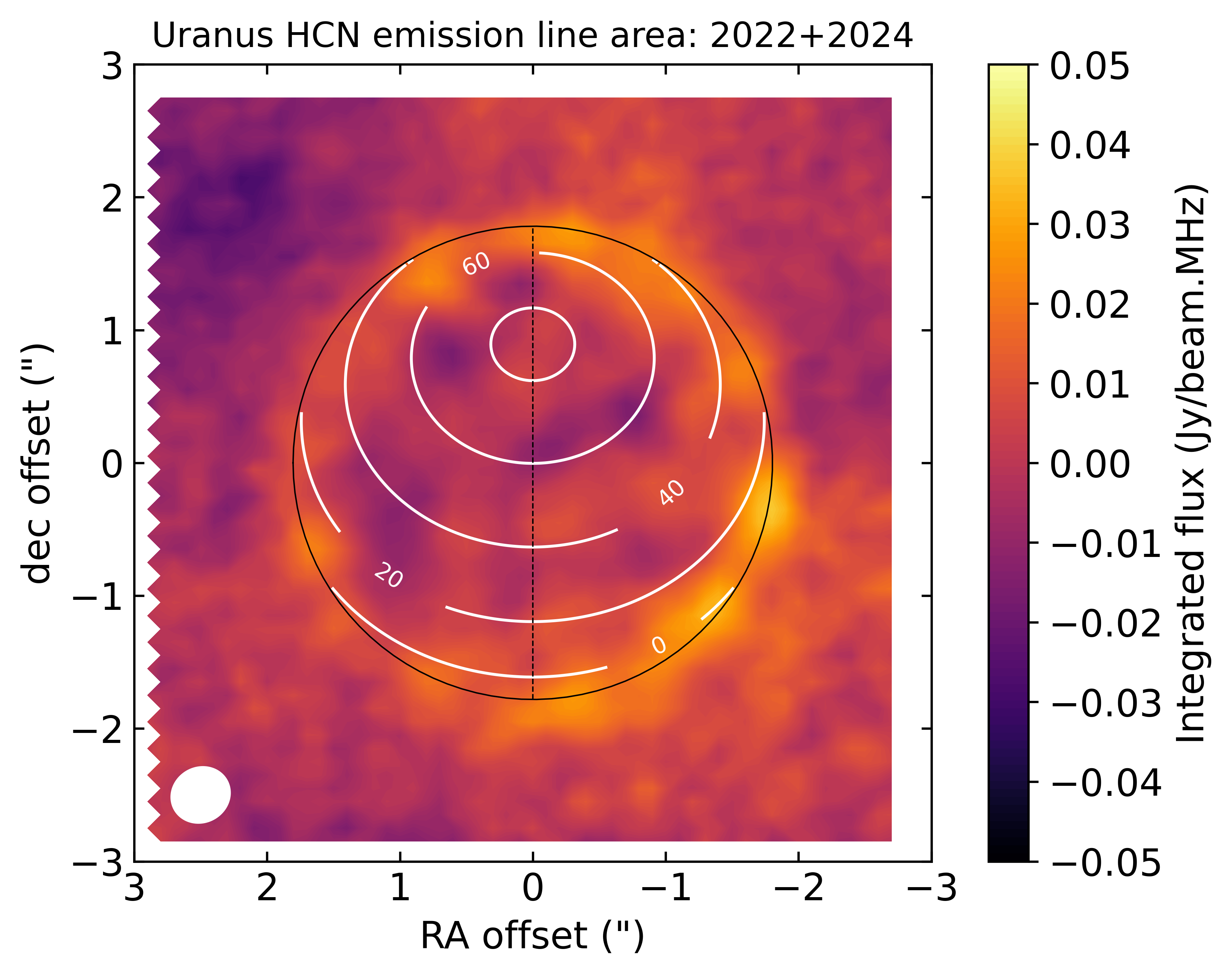}
    \caption{Uranus HCN (J=4-3) emission line area map from the 2022$+$2024 observations. This illustrates the stratospheric emission detected at the limb by limiting the flux integration from -5\kms~to +5\kms. The layout is the same as in \fig{fig:continuum}. }
    \label{fig:linearea_HCN}
  \end{figure}

\begin{figure}[!h]
    \centering
    \includegraphics[width=12cm]{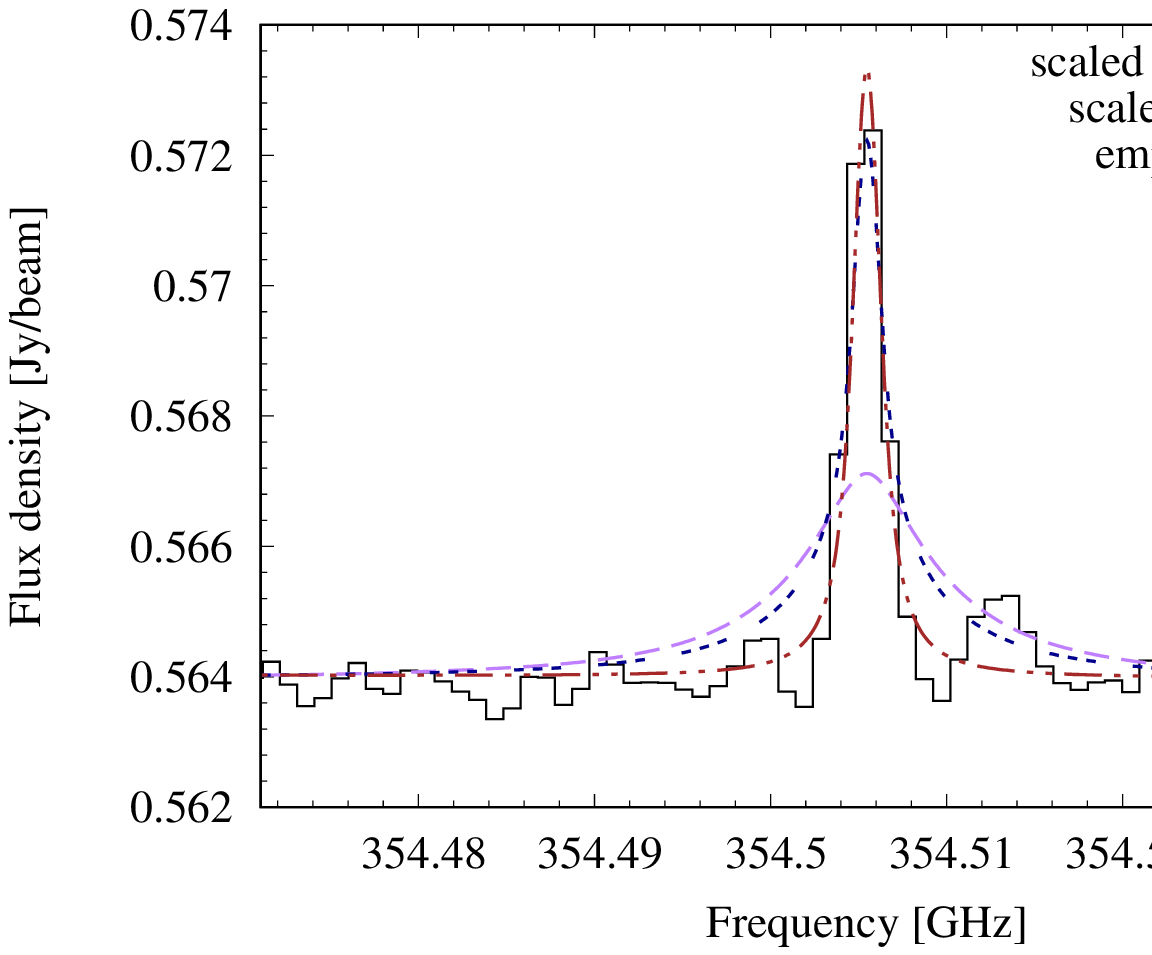}
    \caption{Uranus HCN (J=4-3) emission line after averaging the signal from the 2022$+$2024 cube over all limb positions (see Section \ref{sec:HCN-obs}). The best-fit model is obtained with an HCN vertical profile in which HCN is restricted to pressures lower than 0.2 mbar with a mole fraction of 1.8\dix{-11} (dashed-dotted brown lines). While the model using a profile based on a rescaled CO comet impact profile and accounting for condensation cannot reproduce the data (long-dashed purple lines), using a rescaled CO IDP profile and also accounting for condensation almost matches the observation (triple-dashed dark blue lines). The wings are slightly too broad but remain within the limit of 3-$\sigma$. }
    \label{fig:final_spectrum_HCN}
  \end{figure}

  For the HCN line, we took the 2022$+$2024 data cube to maximize S/N, following the same methodology as for the CO 2022$+$2024 combined spectral image. A narrow emission line is detected at the planetary limb with an S/N ranging from 3 to 8, as shown in \figs{fig:linearea_HCN}{fig:spectra_HCN}. We thus proceeded with line extraction from the limb with the same spatial sampling as for CO, oversampling the beam by a factor of 4. Because those lines are extracted from a cube mixing data from 2022 and 2024 when the geometry was different, we did not aim at deriving any meridional information on the HCN distribution. To increase the S/N and thus improve the sensitivity of the data to the vertical distribution of HCN, we applied the methodology developed by \citet{Lefour2026}. The extracted spectra were first corrected for the beam-convolved planet rotation to realign them on the line rest frequency before averaging them together. The resulting spectrum is shown in \fig{fig:final_spectrum_HCN}. Any latitudinal variability is obviously lost in the process, but the vertical sensitivity is increased reaching a final S/N of 25 over an 8.4 mJy/beam emission line. The final line is narrow, with a full-width at half-maximum (FWHM) of $\sim$2\,MHz. It also implies that the spectral sampling of the line is quite limited, given the native spectral resolution of 1.1\,MHz.

     \begin{table}[!th]
         \caption{Geometry data of Uranus at the time of the ALMA observations.}
         \label{tab:geometry}
         \begin{center}
         \begin{tabular}{lllll}
            \hline
            Date & $\theta_p$ & $\phi$ & $\lambda$ & NP \\
            \hline
            2022-08-19 10:15UT & 3.61 & 61.3 & 320.7 & 269.1 \\
            2022-10-18 06:30UT & 3.76 & 60.2 & 73.6 & 268.3 \\
            2024-07-30 11:30UT & 3.54 & 68.5 & 276.2 & 276.6 \\
            \hline
         \end{tabular}\\
         \end{center}
         Note: $\theta_p$ is the planet equatorial angular diameter (arcsec), $\phi$ is the sub-observer latitude, $\lambda$ is the sub-observer longitude, and NP is the North Pole angle.
     \end{table}

     \begin{table*}[!th]
         \caption{Configuration of the ALMA observations.}
         \label{tab:ALMA-obs}
         \begin{center}
         \begin{tabular}{llllllll}
            \hline
            Date & pwv (mm) & Nb ant. & Config. & Baselines &  $\Delta\nu$ & $\Delta t$ &  Beam \\
            \hline
            2022-08-19 10:15UT & 0.28 & 44 & C-5 & 15.1-1301.6\,m & 1.1 & 2371.4 & 0.232''$\times$0.201'' PA$=$36.19$^\circ$ \\
            2022-10-18 06:30UT & 0.44 & 45 & C-3 & 15.0-457.0\,m & 1.1 & 2583.1 & 0.696''$\times$0.607'' PA$=$61.46$^\circ$ \\\
            2024-07-30 11:30UT & 0.44 & 52  & C-5 & 15.1-1397.8\,m & 1.1 &  2583.0 & 0.220''$\times$0.182'' PA$=$8.66$^\circ$ \\
            Combined-2022 (CO)        &          &       &       &                  & 1.1 &              & 0.465''$\times$0.428'' PA=-215.93$^\circ$ \\ 
            2022$+$2024 (CO)             &        &       &       &                  & 1.1 &              & 0.469''$\times$0.436'' PA=-18.78$^\circ$ \\
            Combined-2022 (HCN)        &          &       &       &                & 1.1 &              & 0.442''$\times$0.413'' PA=-52.94$^\circ$ \\
            2022$+$2024 (HCN)             &        &       &       &                & 1.1 &              & 0.442''$\times$0.403'' PA=-60.72$^\circ$ \\
            \hline
         \end{tabular}\\
         \end{center}
         Note: pwv stands for precipitable water vapor, Nb ant. for the number of antennas, Config. for the array configuration, $\Delta\nu$ for the spectral resolution in MHz, $\Delta t$ for the on-source integration time in seconds, and Beam gives the synthetic beam properties at the frequency of CO for the individual dates, and for both CO and HCN for the combined datasets. PA is the Position Angle of the synthetic beam, which is counted from the northern direction in the trigonometric sense. 
     \end{table*}

  \begin{figure}[!h]
    \centering
    \includegraphics[width=12cm]{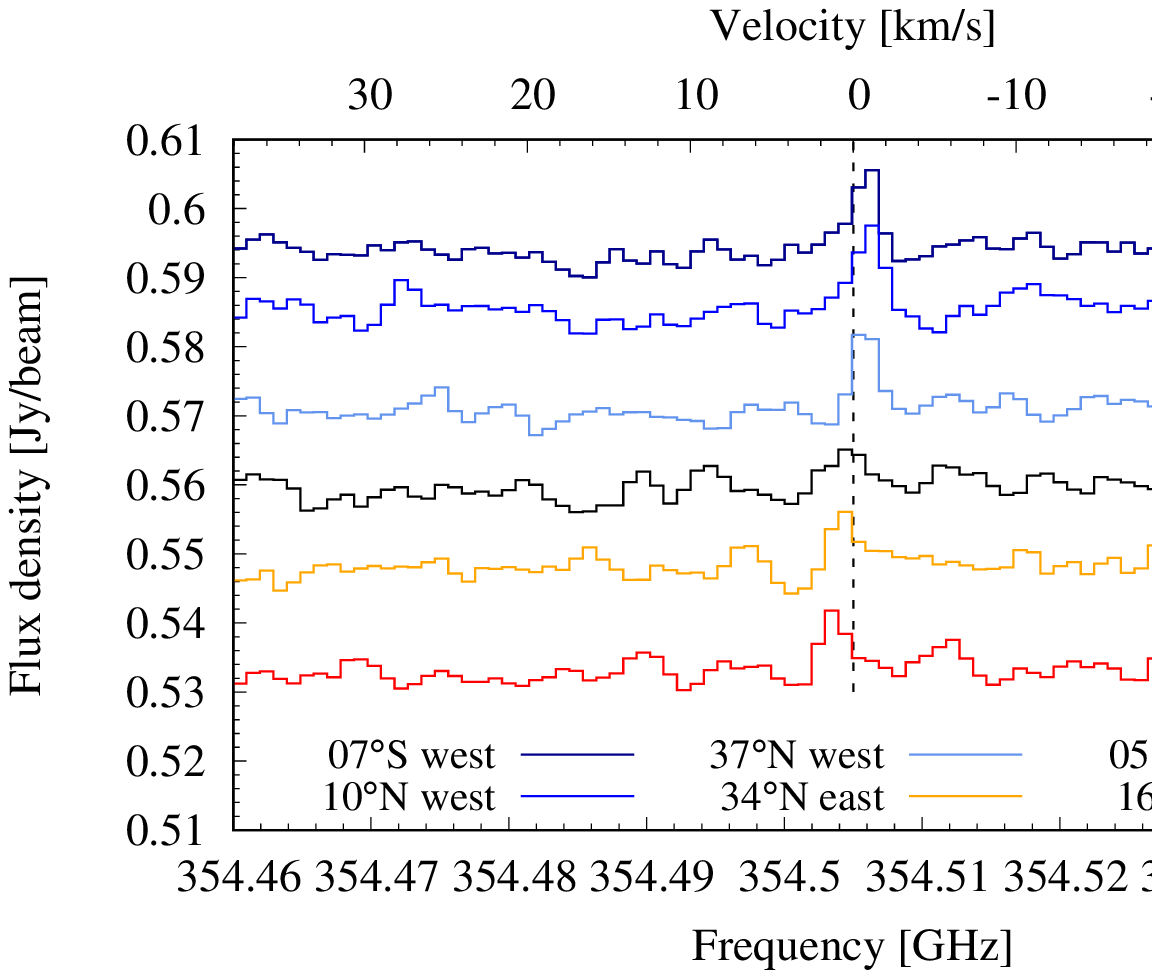}
    \caption{Selection of Uranus HCN (J=4-3) spectra across the planet limb, demonstrating the low-to-moderate S/N detection of the emission line. To avoid confusion, spectra are offset as follows: 07$^\circ$S west (dark-blue line) by $+$0.04\,Jy/beam, 10$^\circ$N west (blue line) by $+$0.02\,Jy/beam, 37$^\circ$N west (light blue line) no offset, 16$^\circ$S east (black line) no offset, 34$^\circ$N east (orange line) by -0.02\,Jy/beam, and 05$^\circ$N east (red line) by -0.03\,Jy/beam. The rest frequency of the HCN (J=4-3) line is indicated by the vertical dashed gray line, to illustrate the line shifts caused by the planet rotation.}
    \label{fig:spectra_HCN}
  \end{figure}

\section{Modeling} \label{sec:Models}
  \subsection{Model inputs}
    \subsubsection{Background composition}
    In our modeling, we accounted for the species that contribute to atmospheric opacity over the observed spectral ranges. In the atmosphere of Uranus, these include H$_2$, He, CH$_4$, through collision-induced absorption (CIA), NH$_3$ through the opacity caused by broad wings of tropospheric lines, and CO and HCN through the J=3-2 and J=4-3 (respectively) line opacities. We used the He abundance derived from Voyager 2 measurements by \citet{Conrath1987}. The tropospheric CH$_4$ mole fraction has been demonstrated to vary with latitude at pressures larger than 1 bar in Uranus and Neptune \citep{Baines1995,Karkoschka2009,Karkoschka2011,Sromovsky2008,Sromovsky2011,Sromovsky2014,Irwin2019b}, with a peak at low latitudes and minima at high latitudes. However, our results are only
sensitive to the CH$_4$ abundance in layers located at $\sim$0.1-1\,bar, because the species only contributes to the continuum through H$_2$-CH$_4$ CIA in the this pressure range. We thus adopted the profile derived by \citet{Lellouch2015a} from combined Herschel and Spitzer observations, in which a deep CH$_4$ mole fraction of 0.032 decreases because of condensation in the upper troposphere following a 75\%~relative humidity from 800\,mbar to 100\,mbar, and then further decreases with a log-log slope from 4.5\dix{-5} at 89\,mbar to 1.6\dix{-9} at 2.5\,mbar, before finally decreasing because of the homopause (at $\sim$0.1\,mbar). Following \citet{dePater1989a}, we also accounted for a low abundance of 6\dix{-6} for upper tropospheric NH$_3$. This species is more abundant in the deeper layers of Uranus and Neptune \citep{Molter2021,Tollefson2021}, but is vastly removed from the gas phase around $\sim$30\,bar, where it is expected to combine with H$_2$S to form an NH$_4$SH cloud. Despite a retrieved deep abundance of $\sim$1.7\dix{-4} in Uranus \citep{Molter2021}, we did not account for this deep component of the NH$_3$ profile, because our calculations are insensitive to it. Even though H$_2$S is present in the upper troposphere of Uranus \citep{Irwin2018}, the spectral lines of this species in the 100-400\,GHz range were not accounted for in the modeling. They essentially alter the continuum at frequencies below 300\,GHz \citep{dePater2023b}, because of their broad wings, and consequently do not impact our analysis around 350\,GHz. The H$_2$ abundance was taken as the complement to unity of the sum of all other abundances.

    \subsubsection{CO and HCN distribution models \label{sec:compo}}
    In what follows, we constrain the CO profile in a two-step approach. We first used the disk center absorption to constrain the CO abundance from $\sim$10\,mbar (lower stratosphere) to $\sim$500\,mbar (upper troposphere), because this spectrum is only sensitive to this deeper component of the CO profile. We then used the limb emission spectra to constrain the CO abundance between 0.1 and 1\,mbar (upper stratosphere), that results from an external source. The final CO profile we derived from this work is then the addition of the internal source profile with the external source one. 
    
    The internal CO was modeled with a constant profile set to the deep CO abundance at the lower boundary of the model atmosphere, i.e., in the troposphere. This profile sharply decreases at the homopause because of molecular diffusion. The homopause level is found at much higher pressures than in other giant planets ($\sim$0.1\,mbar), because of the low eddy diffusion coefficient in the stratosphere of Uranus (see \citealt{Moses2017} for a comparison). We adopted the shape of the profile computed by \citet{Cavalie2014} and rescaled it appropriately to reproduce the disk center absorption.

\begin{figure}[!t]
    \centering
    \includegraphics[width=12cm]{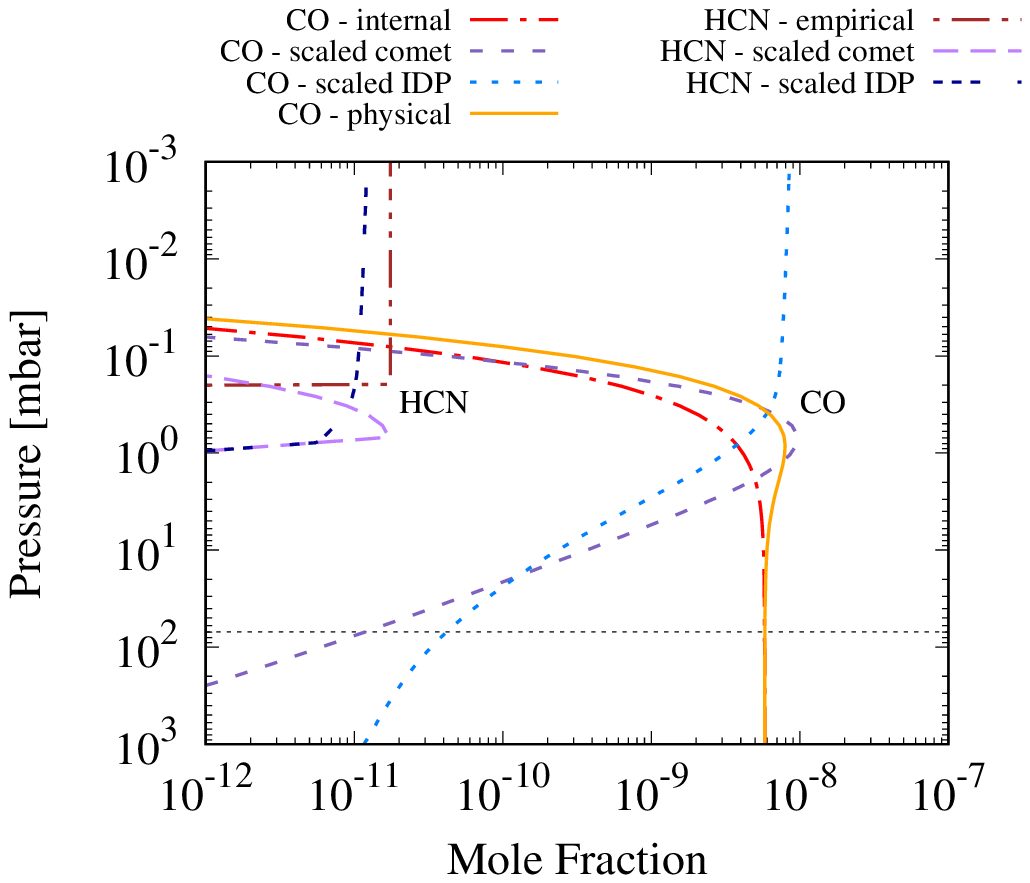}
    \caption{Mole fraction vertical profiles considered in this work for CO and HCN. For CO, we test the profiles based on \citet{Cavalie2014} that reflect: (i) an internal source with 5.8\,ppb (dashed-dotted red line), as determined from the disk center spectrum (see text in Section \ref{sec:Internal}), and (ii) an external source that can be an old comet impact (dashed dark-purple line) or a steady IDP flux (dotted blue line). These profiles are discussed in Sections \ref{sec:Internal} and \ref{sec:Meridional}. The ring source model described in Section \ref{sec:compo} is based on the IDP model and is not shown here. The orange solid line adds an old comet impact onto an internal source, as obtained from a vertical transport model (see details in Section \ref{sec:Combined}). Contrary to the profiles from \citet{Cavalie2014} that were produced with $K_{zz}$$=$1200\Kunit, this dual source profile is obtained with $K_{zz}$$=$2430\Kunit~\citep{Orton2014b,Moses2018}. For HCN, we use an empirical step profile, and rescaled CO comet and IDP models that account for condensation, as presented in Section \ref{sec:HCN}. The layout is the same as in \fig{fig:final_spectrum_HCN}. The tropopause level is indicated with a thin dotted black line. All models correspond to the best-fit models discussed in the paper.}
    \label{fig:fm_profiles}
  \end{figure}

    We defined three external source models for CO:
    \begin{enumerate}
      \item Large comet impact: we took as reference the comet profile derived by \citet{Cavalie2014} using a vertical transport model to fit Herschel-HIFI observations of the CO (J=7-6) line. In their model, it corresponds to a comet with a diameter of 640\,m yielding 50\%~of CO impacting Uranus 300 years ago and depositing its material at pressures lower than 0.1\,mbar. We will discuss in Section \ref{sec:Meridional} the fact that several comet-impact-derived profiles resulting from different combinations of size and impact date could fit the Herschel observations equally well.
      \item IDP source: we also adopted as baseline the vertical profile derived by \citet{Cavalie2014} that results from a uniform input flux of 2.2\dix{5}\Fluxunit.
      \item Local source (icy rings and satellites): we built this source model by combining the IDP profile presented above, scaled as a function of latitude by a gaussian function peaking at the equator and with an FWHM of 25$^\circ$ latitude, following the results obtained by \citet{Cavalie2019} on the latitudinal distribution of H$_2$O in Saturn. We thus have a distribution that peaks at the equator and decreases with increasing latitude. The overall distribution was scaled by a factor of 2.6 to match the CO (J=7-6) line emission observed with Herschel. We note that the Herschel observations were taken in 2011-2012, when the sub-observer latitude was in the range 17$^\circ$N-21$^\circ$N planetocentric latitude.
    \end{enumerate}
    Comet impact and IDP profiles scaled to match the ALMA observations, as discussed in Section \ref{sec:Meridional}, are presented in \fig{fig:fm_profiles}. After the derivation of the internal and external sources, we verified that the internal+external source profile did not alter the quality of the fits to the disk-center absorption line and to the limb emission lines.
  
    The detection of HCN as a narrow emission line in our limb average implies that the observed HCN is located in the upper stratosphere. In a first step, we simply adopted a two-level vertical profile, in which HCN is only present in the upper stratosphere, i.e., at pressures lower than some cut-off pressure that we determine from the observations. In a second step, we rescaled the best-fitting CO profile to determine the CO/HCN ratio. One important point to note is that HCN condenses at around 1\,mbar in the cold Uranus stratosphere \citep{Fray2009}. We therefore corrected the rescaled profile to account for condensation.

    \subsubsection{Atmospheric temperature \label{sec:temps}}
    For what concerns the atmospheric thermal field, we implemented the latitude-dependent field retrieved by \citet{Roman2026} from JWST observations performed on January 8, 2023, i.e., right in-between the ALMA observations presented in this paper. Given the geometry at the time of the observations, temperatures were only retrieved from the equator to the North Pole of the planet. For the southern hemisphere that was only very partially observed with ALMA, we used the equatorial temperatures. The retrieved temperatures have uncertainties of $\pm$0.7\,K near the 1\,bar level, $\pm$1.5\,K at 0.2\,mbar, and $\pm$4\,K near the top of the model (\ten{-4}\,mbar). Temperature profiles taken at various latitudes are relatively similar to the disk-averaged profile constrained from Spitzer observations of 2007 by \citet{Orton2014a}, even if the season had evolved from equinoctial in 2007 to northern spring in 2023 (northern summer solstice will occur in 2030). One noticeable trend though is found at 0.2\,mbar from the equator to 45$^\circ$N with a 6\,K increase in the JWST temperatures.

  \subsection{Radiative transfer model}
  We used the line-by-line radiative transfer model described in \citet{Cavalie2019,Cavalie2026}, adapted to the ellipsoidal shape of Uranus \citep{Lindal1987}. The continuum opacity resulting from collision-induced absorption (CIA) of H$_2$-H$_2$, H$_2$-He and H$_2$-CH$_4$ pairs was accounted for using the work from \citet{Borysow1985}, \citet{Borysow1988}, and \citet{Borysow1986}, respectively. Additional continuum opacity caused by NH$_3$ line broad wings, although relatively negligible, was also included. Spectroscopic parameters were taken from the JPL database\footnote{\url{https://spec.jpl.nasa.gov/}} \citep{Pickett1998}. For NH$_3$, we took the same broadening coefficients as those presented in \citet{Cavalie2026}. For the CO (J=3-2) line, and based on the work of \citet{Dick2009}, we adopted $\gamma=0.065$\unitgamma~and $n=0.54$ at a reference temperature of $T_\mathrm{ref}=296$\,K. For the HCN (J=4-3) line, and based on the work of \citet{Rohart1987}, we took $\gamma=0.143$\unitgamma~and $n=0.75$ at a reference temperature of $T_\mathrm{ref}=300$\,K.

  \subsection{Thermochemical model \label{sec:thermo_model}}
  With the detection of upper tropospheric CO in Uranus, we can constrain the deep O/H ratio of the planet with thermochemical and vertical mixing modeling. We applied the 1D model already extensively presented in \citet{Cavalie2017,Cavalie2023a,Cavalie2024} that uses the chemical scheme of \citet{Venot2020}. At the high temperatures prevailing in the deep tropospheres of giant planets, oxygen is almost exclusively borne by H$_2$O molecules. Thermochemical and transport modeling then enables to link the upper tropospheric CO abundance to the deep H$_2$O abundance by combining vertical mixing with the solving of the following equilibrium equation: H$_2$O+CH$_4$=3H$_2$+CO. The deep H$_2$O abundance then gives directly the planet deep O/H abundance ratio. The deep tropospheric temperature profile was obtained by extrapolating the temperature at 2\,bars following \citet{Leconte2017} to account not only for the wet adiabat but also for the effect of the mean molecular weight gradient. 
  
  While the deep troposphere is in thermochemical equilibrium, owing to the high temperatures, equilibrium is quenched higher up by vertical transport when the decreasing temperature increases the chemical timescale beyond the mixing timescale. It results in fixing the CO abundance to a value orders of magnitude higher than what would be expected from pure thermochemical equilibrium (see e.g., \citealt{Visscher2010}). One key parameter is then the magnitude of vertical mixing, which is parametrized by a vertical eddy diffusion coefficient in 1D models, taken to be constant throughout the troposphere. According to the mixing length theory, its value is related to the internal heat flux of the planet. The recent determination of Uranus' internal heat flux by \citet{Irwin2025} and \citet{Wang2025} only increases the magnitude of vertical mixing by about 30\%~compared to the previously assumed value of 10$^8$\Kunit~\citep{Cavalie2020}. This has a very limited impact on the derived O/H ratio compared to the overall uncertainties on the value of this parameter, as discussed Section \ref{sec:Internal}.

  \subsection{Vertical transport model \label{sec:transport_model}}
  The physical profiles computed for each source by \citet{Cavalie2014} used a stratospheric $K_{zz}$ of 1200\Kunit. However, \citet{Orton2014b} constrained its value to 2430\Kunit, a value later used by \citet{Lellouch2015a} and \citet{Moses2018}, placing the methane homopause at 0.1\,mbar. After deriving the upper tropospheric CO abundance and the best source from the rescaled profiles of \citet{Cavalie2014} for the stratosphere, we used a vertical transport model with $K_{zz}$$=$2430\Kunit~to compute a new and complete vertical profile of CO, including an internal and an external source, to reproduce the complete CO dataset. 
  
  In this model, the internal sources were set as fixed abundances at the lower boundary of the atmosphere. The model was then run until steady state was reached, typically more than 10$^{10}$\,s, so as to have the homopause properly modeled for all species. The external sources were added in a second step and the model parameters representing the HCN and CO sources were optimized to fit the data. IDPs were modeled with a constant flux at the upper boundary of the atmosphere. The model was then run again until steady state was reached, also typically more than 10$^{10}$\,s. Alternately, for the comet impact source, we ran the model for a time corresponding to the time elapsed since the impact and constrained the mole fraction initially deposited above the 0.1\,mbar level.

  \subsection{Wind retrieval model}
  With a moderate nominal spectral resolution of 1.1\,MHz, the CO observations of Uranus presented in this paper were not specifically designed for wind measurements. However, the measurements achieved with similar resolution in Neptune by \citet{Carrion-Gonzalez2023} encouraged us to try the detection of stratospheric winds in Uranus from these data. We thus applied the algorithm developed by \citet{Cavalie2021} and \citet{Benmahi2022}. It fits a parametrized line profile to a spectrum in order to constrain the Doppler shift of the line with respect to its rest frequency using an MCMC scheme. The beam-convolved Doppler shift caused by the planet rotation was then computed with the radiative transfer model and subsequently subtracted to derive the line shift induced solely by the winds. This process was then repeated for each spectrum of the dataset. We applied the algorithm only to spectra extracted from the combined-2022 CO and HCN spectral cubes at the location of the planet limb. Given the relatively large synthetic beam size compared to the atmosphere apparent thickness, we defined the limb position at the 1-bar pressure level. The pressure level at which the winds are probed were constrained by modeling the vertical contribution of these spectral emissions using our radiative transfer model \citep{Cavalie2021}.

\section{Internal CO} \label{sec:Internal}
In this section, we use the 2022$+$2024 spectral cube obtained from the combination of the three Uranus observations recorded between August 2022 and July 2024 to derive the upper tropospheric CO abundance. The reason for the combination of all the available data, despite the different geometries of the planet between 2022 and 2024, is the optimization of the signal at the disk center. The derivation of the upper tropospheric CO abundance from the disk-center spectrum in this work then pertains to latitudes around the sub-Earth point (between 60\degre N and 70\degre N planetocentric).

As baseline, we adopted the shape of the internal source profile of \citet{Cavalie2014}. It essentially translates into a vertically constant mole fraction throughout the troposphere and the lower stratosphere. Because of the low homopause of Uranus \citep{Moses2018}, the CO abundance then sharply decreases at pressures lower than $\sim$1\,mbar, in this internal source profile. By rescaling this profile with a constant factor, we can explore the range of upper tropospheric CO abundances that enables to reproduce the disk-center spectrum. 
  
\begin{figure}[!t]
  \centering
  \includegraphics[width=12cm]{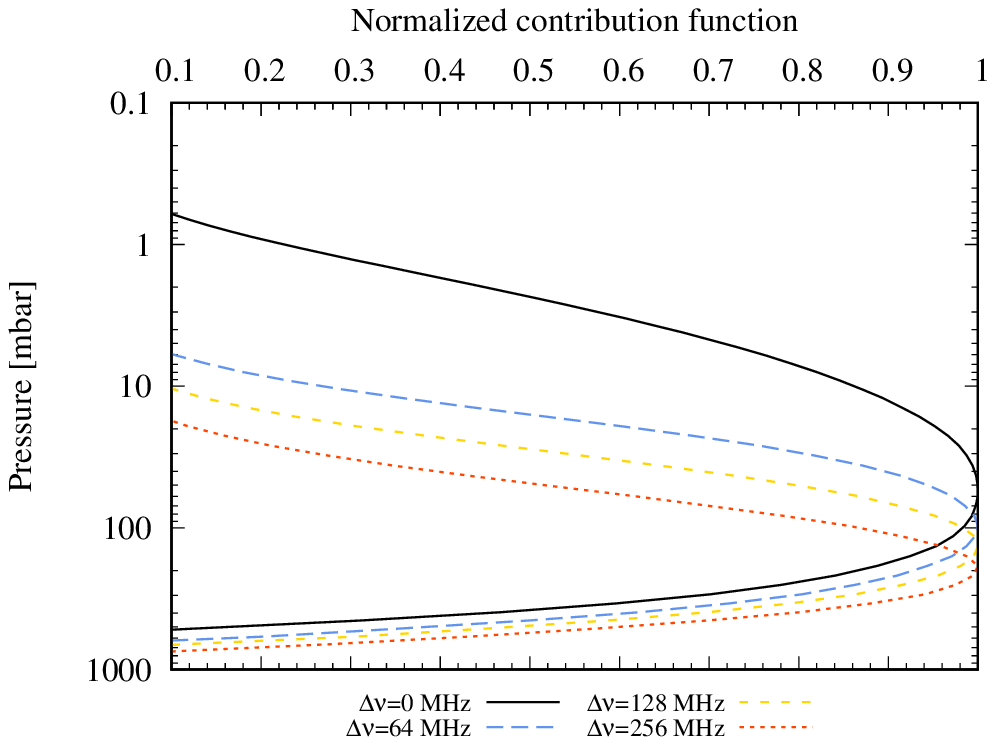}
  \caption{Normalized contribution functions at the line center for the CO (J=3-2) line observed with ALMA in 2022 (solid black line), accounting for the internal source only. Contribution functions are also presented at 64\,MHz (long-dashed blue line), 128\,MHz (short-dashed yellow line), and 256\,MHz (dotted red line), frequency offsets with respect to the line center. They are all computed for the disk center pointing and for a spectral resolution of 8.8\,MHz. The broad wings thus probe the upper troposphere ($p$$>$70\,mbar). }
  \label{fig:contribution}
\end{figure}

We find a difference of $\sim$6\%~between the modeled and observed flux densities in the far wings of the CO line. With up to 5\%~flux calibration uncertainties, the remaining difference may result from 1-2\,K differences at the 1\,bar level between our assumed temperatures (retrieved from JWST observations in January 2023, see Section \ref{sec:temps}) and the actual ones during the observations, as a 1\,K difference already translates into 1\%~difference on the continuum. The fitting of the observations is then achieved by rescaling the model flux density at 345\,GHz to that of the observations, i.e., far enough from the core of the absorption not to influence the derivation of the upper tropospheric CO abundance. The best fit is obtained after $\chi^2$ minimization with 5.8$\pm$0.3\,ppb of upper tropospheric CO. The uncertainty range accounts for the spectral noise and the uncertainties on the temperature. The quality of the fit for the nominal abundance is very good with a $\chi^2/N$ of 1.01 for the N spectral data points comprised between 345.4 and 345.95\,GHz. None of the external source models discussed in the next section can fit this observation. We also tested a ``rock giant '' model similar to those proposed by \citet{Teanby2020}, in which the planet interior is essentially devoid of ices, and only external sources exist and contaminate the atmosphere down to the tropopause of the planet (at 70\,mbar in Uranus). Such a model, with 8\dix{-8} of CO above the tropopause, fails at reproducing the slope of the broad wings of the line, because of a lack of CO below the tropopause in that model. Decreasing the atmospheric temperature in the layers around the tropopause by an unreasonable 5\,K does not help reconcile the rock giant model with the data any better. All these results are illustrated in \fig{fig:CO-disk-center}. Those results combined with the contribution functions at various frequencies from the line center, and displayed in \fig{fig:contribution}, demonstrate that we detect the upper tropospheric component of the CO profile.
  
From the observation of CO fluorescence emission at 4.7\,$\mu$m, \citet{Encrenaz2004} derived an upper limit of 2\dix{-8}~on tropospheric CO that would be expected from an internal source, and favored a case in which all the CO was restricted to the stratosphere with a mole fraction of 3\dix{-8}. \citet{Teanby2013} derived an upper limit of the CO tropospheric abundance of 2.1\,ppb. The tropospheric CO abundance we derive in this work is nominally inconsistent with this number. There are two reasons that can explain this apparent inconsistency: (i) the two datasets do not probe the same latitudinal ranges: Uranus was just a few years past equinox at the time of the Herschel observations of \citet{Teanby2013} with a sub-Earth latitude of 10\degre N, while the ALMA observations were recorded when Uranus had a sub-Earth latitude comprised between 60\degre N and 70\degre N. It remains to be explored whether or not these differences in the seasons and latitudes (and thus in upper tropospheric temperatures) probed by the observations can explain the inconsistency between the upper limit of \citet{Teanby2013} and our detection of 5.8$\pm$0.3\,ppb\footnote{\citet{Visscher2010} and \citet{Wang2015} predict larger $K_{zz}$ at the equator compared to high latitudes (on Jupiter and Saturn), so one might expect more quenched CO at low latitudes than at high latitudes if deep the H$_2$O is uniform. Also, given Juno's inferred equatorial plumes of NH$_3$ and H$_2$O on Jupiter \citep{Bolton2017,Li2020}, and the equatorially peaked CH$_4$ distribution on Uranus \citep{Sromovsky2011,Sromovsky2014}, one might expect more H$_2$O at the equator than high latitudes if H$_2$O is not uniform, leading again to more CO at the equator than at high latitudes.}; (ii) underestimated limitations in bandpass quality in the Herschel/SPIRE dataset that led \citet{Teanby2013} to underestimate their upper limit. 
  
\begin{figure}[!t]
  \centering
  \includegraphics[width=14cm]{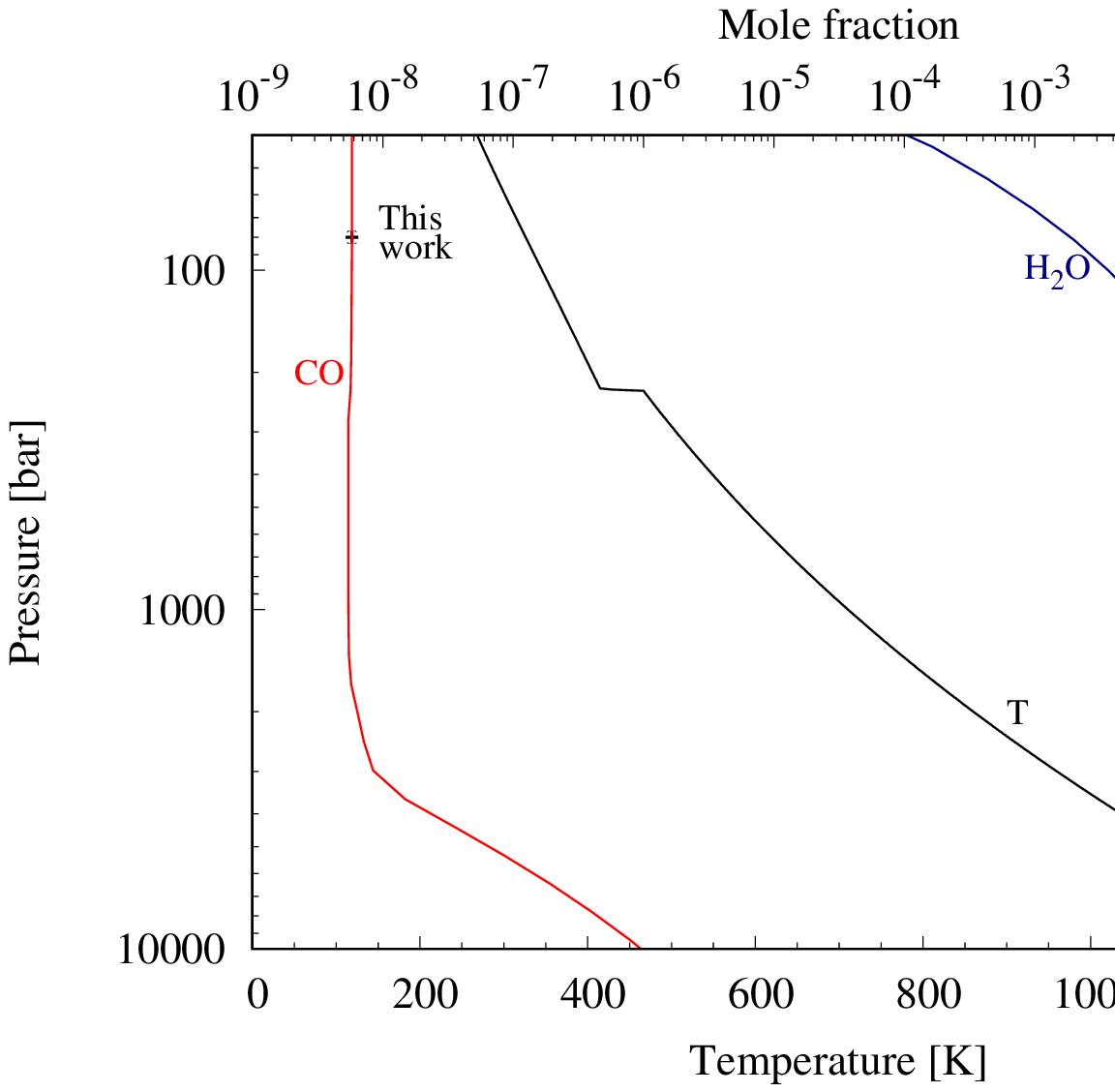}
  \caption{Vertical profile derived from the thermochemical simulations that reproduces the 5.8\,ppb of upper tropospheric CO abundance observed with ALMA. Because the CO profile is constant from $\sim$200\,bar to the upper troposphere, the upper tropospheric CO abundance measured in this work (with its error bar) is indicated at 80\,bar error bar (in black), even though it was made between 0.5 and 0.1\,bar (see \fig{fig:contribution}). The H$_2$, He, CH$_4$, H$_2$O, and CO profiles are plotted in light blue, yellow, green, dark blue, and red. The temperature profile, as obtained with the thermochemical model (see Sect. \ref{sec:thermo_model}, is shown in black. The rapid increase of temperature with depth at $\sim$200\,bar is caused by the mean molecular mass gradient resulting from the H$_2$O condensation. In this model, a vertically constant eddy mixing coefficient of 10$^8$\Kunit~was assumed. }
  \label{fig:abondances_tropo}
\end{figure}

The broad wings of the line probe the upper troposphere up to pressures of a few hundred millibars. It is thus the first unambiguous detection of a tropospheric component of the CO profile in the atmosphere of Uranus. The external source cannot provide such amounts of CO at the deepest level probed by the line (a few 100\,mbar) \citep{Cavalie2014}. We consequently attribute this component of the CO profile to a deep source coming from the interior of the planet. The thermochemical model reproduces the upper tropospheric CO abundance of 5.8$\pm$0.3\,ppb with a deep O/H ratio of 52$\pm$2 times the protosolar value of \citet{Lodders2021}, when assuming 10$^8$\Kunit~for vertical mixing (see \fig{fig:abondances_tropo}). It should be noted that the magnitude of vertical mixing remains uncertain, at least within an order of magnitude, which further expands the error bar on the O/H ratio to 52$^{+30}_{-20}$ for $K_{zz}$$=$\ten{7}-\ten{9}\Kunit. The C/H ratio required in the nominal model to both produce the right amount of CH$_4$ and CO is 58 times the protosolar value, making the C/O ratio of Uranus close to 0.6, i.e., close to the protosolar value of 0.55 \citep{Lodders2021}. However, other sources of uncertainties on the O/H ratio resulting from latitude-dependency of $K_{zz}$ at depth \citep{Wang2015}, the deep tropospheric temperature extrapolation \citep{Cavalie2020}, the chemical scheme \citep{Moses2014,Venot2019,Venot2020}, and chemical mixing barriers \citep{Clement2024,Hyder2025}, may further expand the uncertainty range, especially toward higher O/H. These other sources of uncertainties are quantified more thoroughly in a companion paper by \citet{Briand2026}. If we compare to Neptune, the O/H value we derive for Uranus is lower than that the current estimate for Neptune (200 times protosolar, \citealt{Cavalie2024}), but it nonetheless seems to make Uranus more of an ice giant than a rock giant as proposed by \citet{Teanby2020}. These differences between the two planets may result from different formation locations and pathways for the two planets. This is also addressed in \citet{Briand2026}.

\section{Stratospheric CO meridional distribution} \label{sec:Meridional}
\begin{figure}[!t]
  \centering
  \includegraphics[width=12cm]{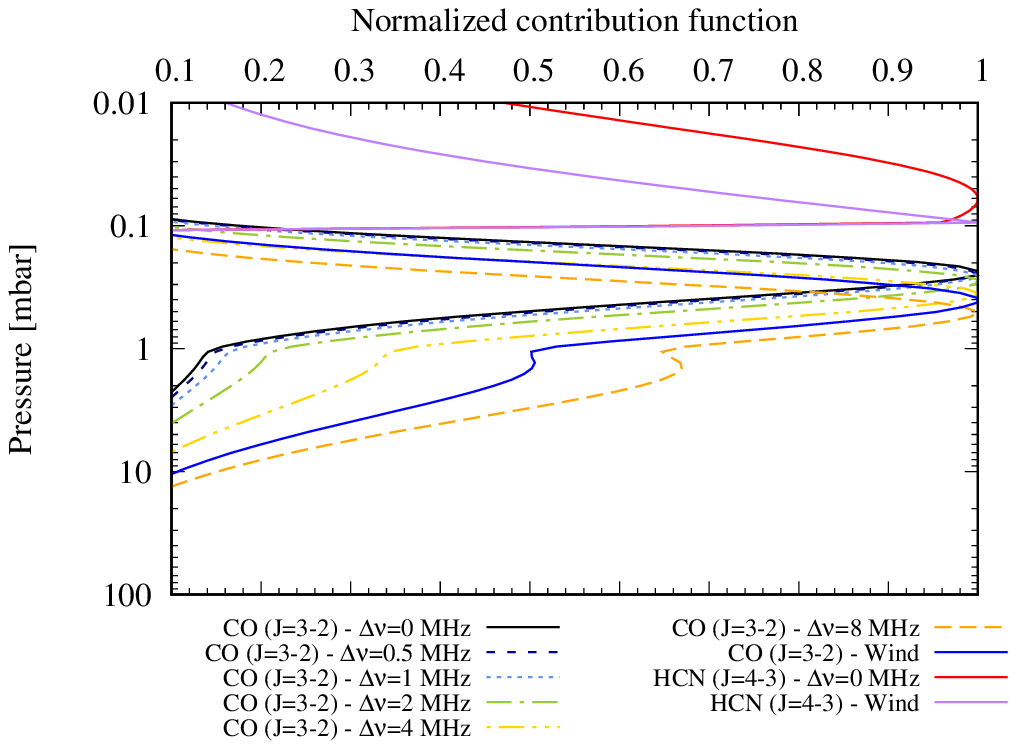}
  \caption{Normalized contribution functions at the center for the CO (J=3-2) line observed with ALMA in 2022 (solid black line), accounting for the comet source and the internal source. Contribution functions are also presented at 0.5\,MHz (dashed dark blue line), 1\,MHz (short-dashed light blue line), 2\,MHz (dashed-dotted green line), 4\,MHz (dashed-double-dotted yellow line), and 8\,MHz (long-dashed orange line), frequency offsets with respect to the line center. The line thus probes the upper stratosphere. The CO wind contribution function is shown with a solid blue line. The normalized contribution at the line center of the HCN (J=4-3) is plotted with a solid red line. The HCN wind contribution function is displayed with a solid purple line. All contribution functions are computed for a limb pointing and for a spectral resolution of 1.1\,MHz. }
  \label{fig:contribution_limb}
\end{figure}

\begin{figure*}[!t]
  \centering
  \includegraphics[width=10cm]{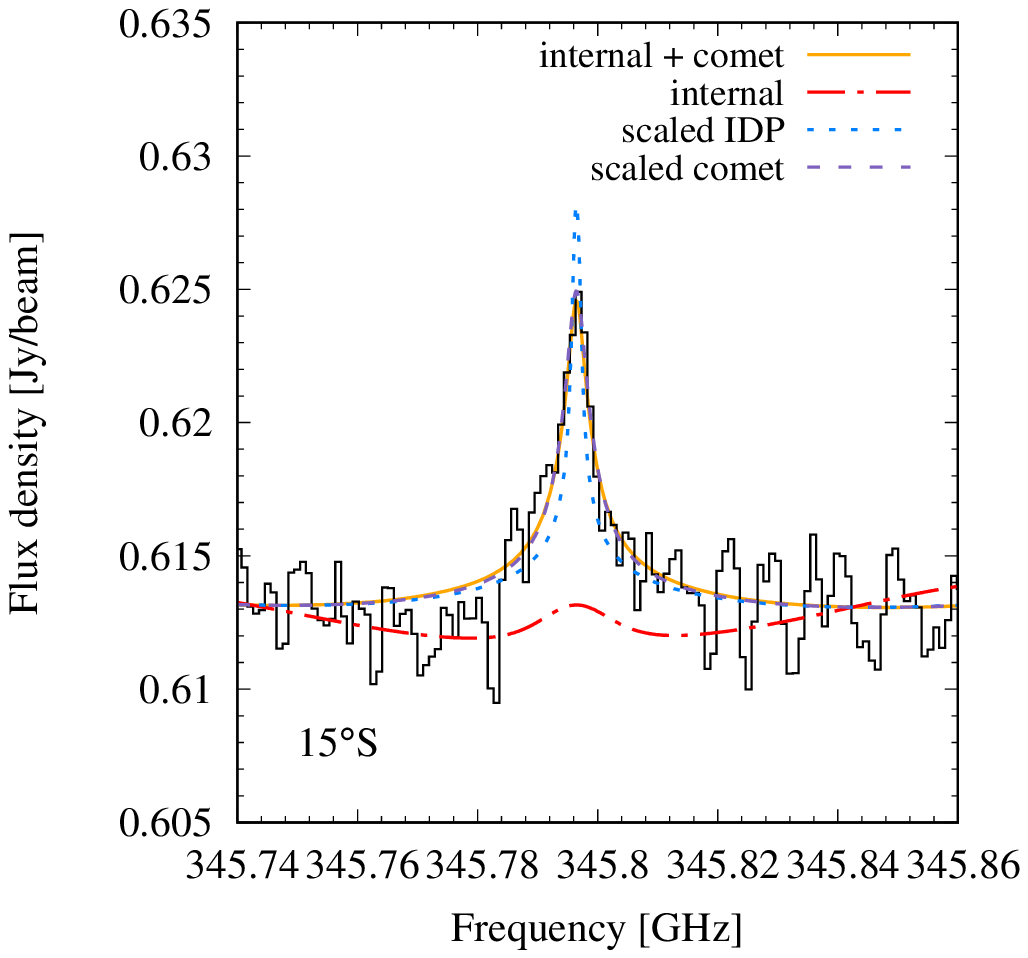}
  \includegraphics[width=12cm]{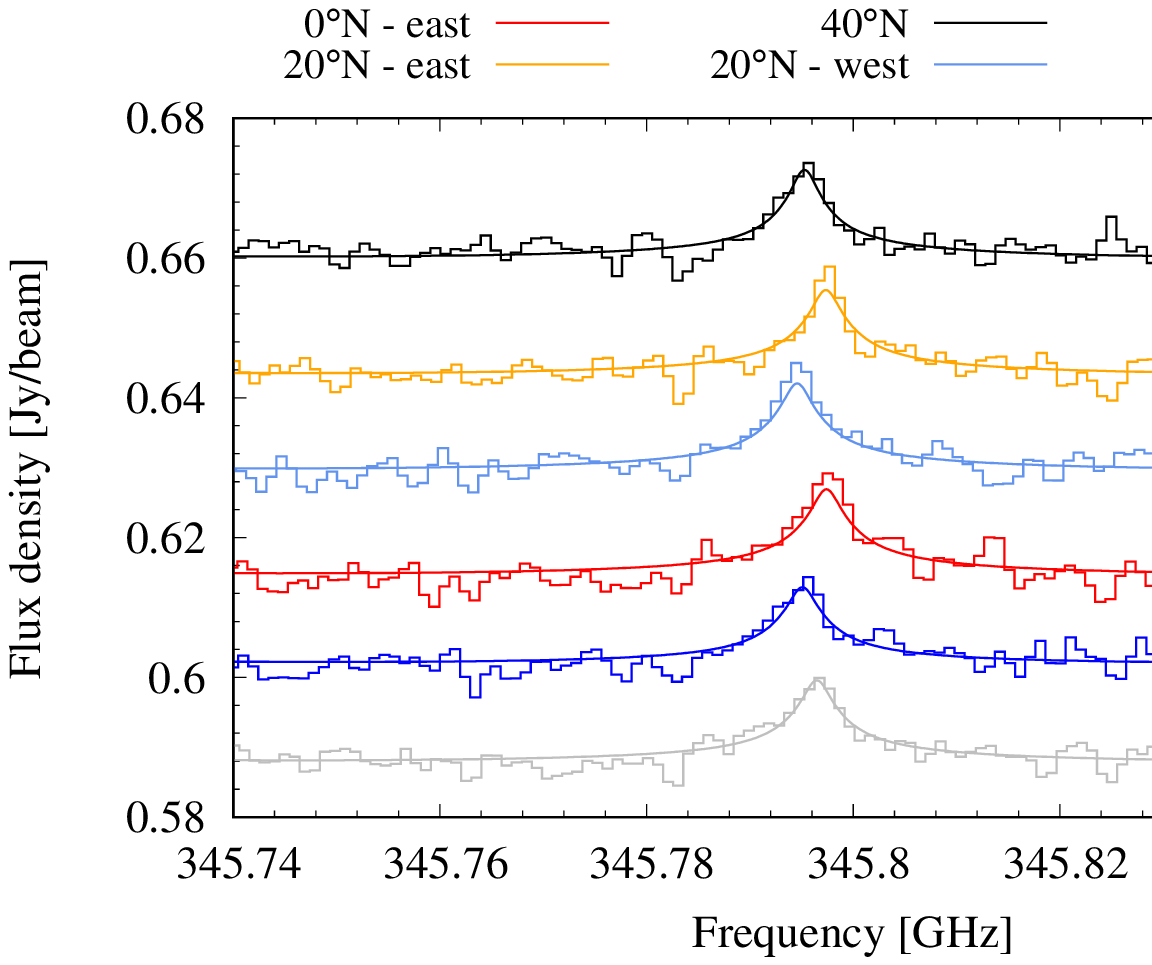}
  \caption{Example of fits to the CO (J=3-2) emission line from the combined-2022 data at various limb positions. (Top) Fits obtained at 15$^\circ$S comparing the various source models: the rescaled comet impact profile (dashed dark-purple lines), the IDP profile (blue dotted lines), and the internal source only (red dashed-dotted line). Finally, the fit obtained with a physical profile produced with a transport model that includes an internal source and a comet impact (see details in Sect. \ref{sec:Combined}) is shown with solid orange line. The latter and the rescaled comet impact model are barely distinguishable from one another. (Bottom) Fits obtained with the dual source model for various latitudes. The spectra are Doppler shifted due to the planet rotation. Some spectra have been shifted in flux for clarity (15$^\circ$S: -0.025\,Jy/beam; 0$^\circ$N west: -0.01\,Jy/beam; 20$^\circ$N west: +0.02\,Jy/beam; 20$^\circ$N east: +0.03\,Jy/beam; 40$^\circ$N: +0.05\,Jy/beam). }
  \label{fig:spectres_CO_comete}
\end{figure*}

The line emission of CO (J=3-2) at the planet limb cannot be reproduced by the sole internal source of CO derived in the previous section (see example at 15$^\circ$S in \fig{fig:spectres_CO_comete} left). The addition of an external source is then required, as shown by the levels probed by the line in the contribution functions (see \fig{fig:contribution_limb}). 

The comet impact model produces fits to the 2022 data with an average $\chi^2/N$=1.0 over all the limb pointings, when the two other sources have $\chi^2/N$ of 1.7, where $N$ is the number of spectral points taken around the line peak ($\pm$5). Example of fits obtained with the comet impact and IDP models are shown in \fig{fig:spectres_CO_comete}. The better fits obtained with the comet impact model come from the fact that the CO profile in this model (which is not a unique solution, as described in \citealt{Cavalie2014}) has a deeper maximum abundance compared to the steady source models. The initial delivery of the cometary material down to the 0.1\,mbar pressure level allows some of the material to penetrate deep enough in the stratosphere, as opposed to the steady source models (IDP and local source), in which CO remains restricted to the upper stratosphere because of the too sluggish vertical eddy mixing \citep{Orton2014a}. Consequently, the comet impact model produces lines that are broader than in the other models, with the appropriate optical depth at the line center ($\tau=$0.16), and thus better fits the observations. 

The average scaling factors for all models are: 0.77$\pm$0.06 for the comet impact, 0.74$\pm0.10$ for IDP, and 0.34$\pm0.06$ for the local source. This enables us to discard the local source model for the parameters we adopted (see Section \ref{sec:compo}), because such a low scaling factor makes this result incompatible with the Herschel observations of \citet{Cavalie2014} for which, by definition, the scaling factor would be unity. This is caused by the fact that the 2022 data presents mostly low-to-mid latitudes at the limb, whereas the Herschel data from 2011-2012 presented a much broader range of latitudes. The ALMA data is thus sensitive to latitudes where the CO abundance is maximum in the local source model. The CO abundance in the local source model actually exceeds that of the IDP model at low-to-mid latitudes, because of the overall 2.6 scaling factor needed to match the Herschel disk-averaged data (see Sect. \ref{sec:compo}). To match the ALMA data, the local source model must be scaled back down by the factor of 0.34 that we derive here. Also, such high temporal or seasonal variability over only $\sim$10 years is not expected given the long chemical and transport timescales involved in the stratosphere of Uranus \citep{Moses2018}. The 2024 data also favors the comet impact model, yet in a less pronounced way, owing to the more limited S/N of the data. The external source scaling factors, as derived from the 2024 data, are fully compatible with those derived from the 2022 data, with 0.81$\pm$0.08 for the comet impact model, 0.81$\pm$0.09 for the IDP model, and 0.33$\pm$0.04 for the local source model. The difference between the CO abundances derived from the Herschel and the ALMA data with the comet impact model is $\sim$20\%. This remains within error bars, given the 15\%~uncertainties in the \citet{Cavalie2014} results, combined to the 8\%~relative uncertainties derived above on the model scaling factors. In the case of the comet impact model, the 10-year time difference between the Herschel and ALMA measurements also adds a few percent change, because of the slow evolution of the vertical profile.

\begin{figure}[!t]
  \centering
  \includegraphics[width=12cm]{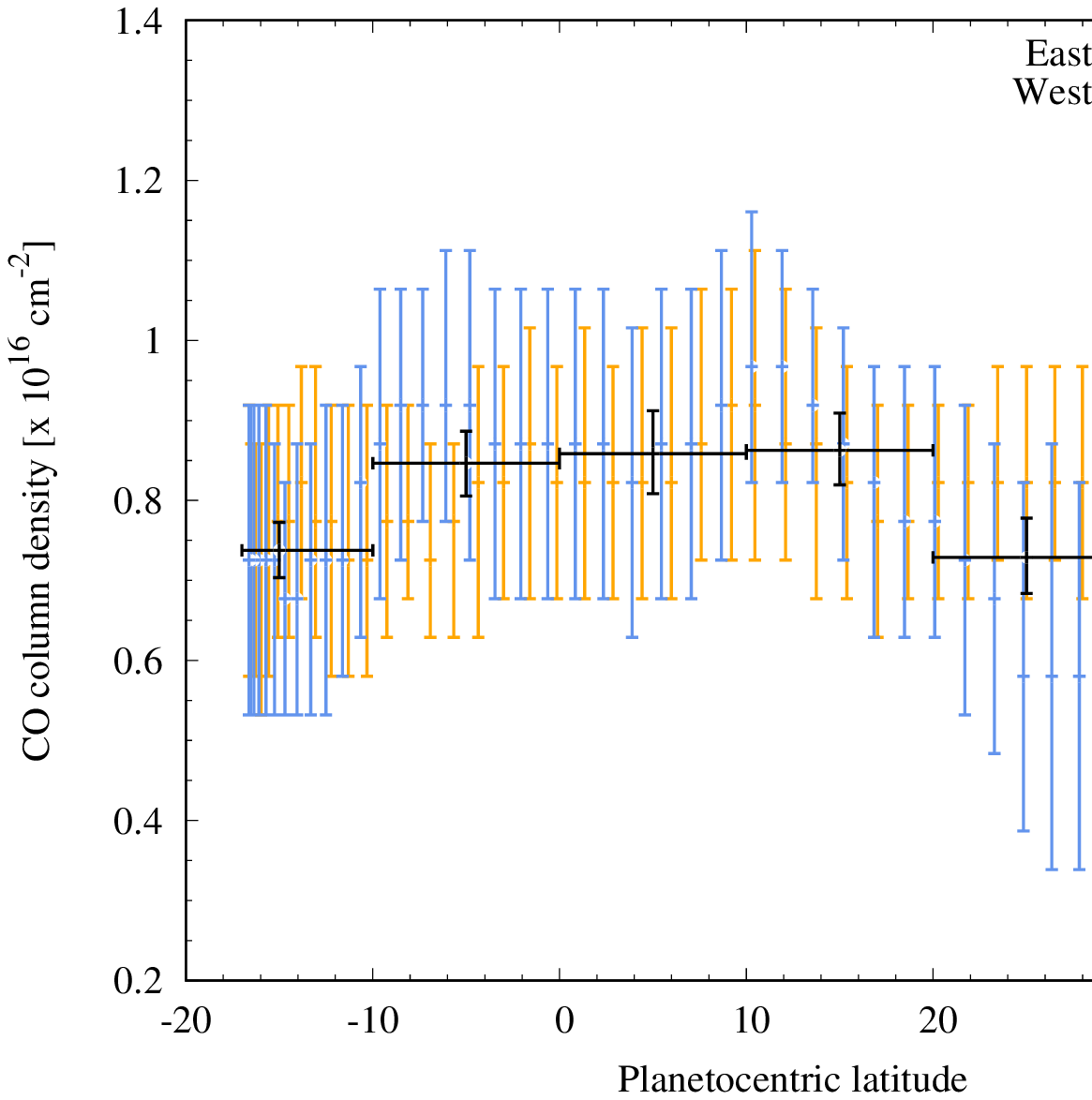}
  \caption{Column density of CO (in cm$^{-2}$) from the external component only as a function of planetocentric latitude, as derived from scaling the comet profile at each observed limb position from the 2022 dataset. The average column is (7.5$\pm$0.6)\dix{15}\Coldens. The same data, after averaging the two limbs and binning them in 10$^\circ$ latitude bins, are shown in black. }
  \label{fig:CO_column_comet}
\end{figure}

When integrating vertically the CO profiles rescaled at each limb position, one can derive the meridional distribution of CO for the range of latitudes covered by the limb observations. This is displayed in \fig{fig:CO_column_comet} and shows that the CO distribution is uniform within error bars with an average column of (7.5$\pm$0.6)\dix{15}\Coldens, in agreement the interpretation of the recent JWST mapping observations of CO \citep{Roman2026}. This column is computed for the contribution of the comet impact alone, and does not count the internal source component of the CO profile. Averaging the results in 10$^\circ$ latitude bins may hint at an equatorially peaked meridional distribution, but this would need to be confirmed by more sensitive measurements. 

This rather uniform distribution can only be compatible with a comet impact if the impact is old enough for the material to have the required time to spread over the whole planet. In our modeling, we took the profile computed by \citet{Cavalie2014} for a 300-years old impact. In Jupiter, horizontal eddy mixing at the 0.1-1\,mbar level was constrained from the temporal evolution of spatial distribution of species deposited by the Shoemaker-Levy 9 comet in 1994, such as CO. Both \citet{Lellouch2002} and \citet{Moreno2003} found it to be 10$^6$-10$^7$ times stronger than vertical mixing at the same pressure level (see \citealt{Hue2018} for a more detailed discussion on this point). If we assume we can similarly scale the vertical mixing found in Uranus by \citet{Orton2014a} to estimate horizontal mixing, $K_{yy}$ would then be on the order of 10$^9$-10$^{10}$\Kunit. Taking the equator-to-pole distance ($\sim$40000\,km) as an upper limit to the representative length $L$ for a full contamination of the planet, the corresponding diffusion time is then $t$$<$$L^2$/$K_{yy}$$\sim$50-500 years, compatible with our initial assumption of 300 years and the possible small meridional variations in the CO upper stratospheric column. It is noteworthy that older impacts would have required a larger comet than the one assumed in \citet{Cavalie2014}, also making them less probable events \citep{Zahnle2003}. They would have produced slightly broader CO lines that would have remained compatible with the data. The main caveat to this conclusion comes from the limited range of latitudes probed by these limb observations (roughly from 15$^\circ$S to 40$^\circ$N). Observations covering a broader range of latitudes will be required in the future to confirm these findings. However, this may not happen before the end of the next decade, given the slow evolution of the observation geometry.

\section{A dual source of CO}\label{sec:Combined}
At this stage, we needed to verify that a physical profile that accounts for an internal source and a comet source can match the whole dataset. We thus ran the vertical transport model of \citet{Cavalie2014} briefly described in Section \ref{sec:transport_model}. We note that while the profiles used in the previous section were obtained by \citet{Cavalie2014} with the same transport model, they used a vertical eddy mixing of 1200\Kunit. Here instead, we used the more recent derivation from \citet{Orton2014b}, confirmed by \citet{Moses2018} that gives 2430\Kunit.

We find that including both an internal source with a tropospheric mole fraction of 5.8\dix{-9} and a comet impact occurring 300 years ago and depositing a 2.4\dix{-7} mole fraction at pressures lower than 0.1\,mbar enables fitting satisfactorily the whole dataset. The resulting vertical profile is displayed in \fig{fig:fm_profiles} and examples of fits are shown in \fig{fig:spectres_CO_comete}. Fits are usually barely distinguishable from the ones obtained by rescaling the comet profile from \citet{Cavalie2014}.

With these comet profile properties, we estimate the mass of CO delivered by the impact to be 2.7\dix{13}\,g. We can then update the comet impact probability. Following the assumptions made by \citet{Lellouch2005} and followed by \citet{Cavalie2014} of a comet with 0.5\,g cm$^{-3}$ density yielding 50\%~of CO upon impact, we then find that the diameter of the comet is 550\,m (i.e., $\sim$4 times smaller than SL9, according to \citealt{Solem1994}), with an unchanged impact probability of once every $\sim$500\,years (within a factor of 6; \citealt{Zahnle2003}).

\section{HCN stratospheric abundance}\label{sec:HCN}
At the observed spectral resolution of 1.1\,MHz, the HCN (J=4-3) line is only a few spectral channels across. Therefore, we adopted a simple empirical shape for the HCN vertical profile that we used in our spectral modeling. In this profile, the HCN mole fraction was set to a constant value above a pressure cutoff level $p_0$. For $p$$>$$p_0$, the HCN abundance was set to zero. Because the HCN line is relatively narrow (FWHM$\sim$2\,MHz), it implies that the bulk of HCN is located in the upper stratosphere (see also contribution function in \fig{fig:contribution_limb}). With this kind of step profile, we find acceptable fits with $p_0$$=$0.2\,mbar and a mole fraction of (1.8$\pm$0.2)\dix{-11}. This corresponds to a column density of (1.1$\pm$0.1)\dix{12}\Coldens. Obviously, the line is optically very thin, with $\tau$=0.016 at the line center. 

Such a low cutoff pressure makes internally sourced N$_2$ as the source of the detected HCN unlikely at first. However, assuming protosolar nitrogen in the deep troposphere \citep{Molter2021}, our thermochemical model predicts an upper tropospheric abundance of N$_2$ of 4\dix{-8}. Because of the low homopause of Uranus, the N$_2$ abundance decreases to 5\dix{-9} at 0.2\,mbar, an abundance $\sim$300 times higher than that of the HCN detected with ALMA. HCN can be produced from the photolysis of N$_2$ followed by reactions of atomic N with hydrocarbons such as CH$_3$, as modeled by \citet{Moses2023} in Saturn's stratosphere. They showed that this nitrogen-hydrocarbon photochemistry results in a N$_2$/HCN ratio of 200-400 across the stratosphere. Assuming the same chemical efficiency in the production of HCN in the upper stratosphere of Uranus from internally sourced N$_2$ makes N$_2$ a possible source of the observed HCN. The HCN profile would nonetheless follow the N$_2$ profile from its condensation level up to its sharp decrease at the homopause. A loss mechanism (such as adsorption onto aerosols) would then be required at $p$$>$$p_0$ for the vertical profile to be compliant with the obsevations.

We can compute the ratio between the stratospheric column densities of CO and HCN. We find that CO/HCN$=$6800$\pm$1200, when considering the rescaled comet impact profile for CO and the step profile for HCN from above. This very large value could be taken as evidence for a common comet impact origin for CO and HCN, in a similar way as in Jupiter \citep{Moreno2003} and Neptune \citep{Lellouch2005}, where large CO/HCN ratios were measured. However in this case, CO and HCN should share a similar vertical distribution, as in the low-to-mid latitudes of Jupiter \citep{Cavalie2023b}. 

Using the shape of the CO comet impact profile, accounting for the HCN condensation that occurs at 0.6-0.7\,mbar \citep{Fray2009}, and rescaling it down to try to fit the faint HCN emission, produces model lines that are much too broad (FWHM$>$5\,MHz; see example in \fig{fig:final_spectrum_HCN}). This logically results from the difference in pressure found between the $p_0$ level and the level at which the CO profile peaks (i.e., close to the HCN condensation level; see \fig{fig:fm_profiles}). The HCN and CO profiles are thus very different in the stratosphere of Uranus, possibly pointing to: (i) a different source, (ii) the role of aerosols found in the stratosphere of Uranus \citep{deKleer2015,Roman2018} in scavenging HCN at pressures larger than a few 0.1\,mbar, similarly to what happens in Jupiter's polar regions also at pressures larger than 0.1\,mbar \citep{Cavalie2023b}; (iii) or a combination of both. To try the first hypothesis, we used the shape of the CO IDP profile. Rescaling it down by a factor of 700, and still accounting for condensation, provides us with a better fit to the data. The line wings are is still slightly too broad, especially at the foot of the line, but remain at the limit of the 3-$\sigma$ (see \fig{fig:final_spectrum_HCN}). Such a model then corresponds to an influx of HCN of (2.4$\pm$0.3)\dix{2}\Fluxunit. \citet{Teanby2022} derived an influx of H$_2$O of 8.3$^{+4.0}_{-0.9}$\Fluxunit. The ratio between the two is then (0.3$^{+0.2}_{-0.1}$)\%, and is compatible with IDP that would be produced by comet activity \citep{Bockelee2017}. 

The source of the observed upper stratospheric HCN thus remains elusive for now. Its determination requires more complete modeling backed up by higher sensitivity and spectral resolution data to better constrain the HCN vertical profile.

\section{Stratospheric winds} \label{sec:Winds}
High spatial resolution observations containing stratospheric spectral signatures can, in principle, be exploited further to constrain stratospheric wind speeds from the measurement of the Doppler shift induced on the lines by the winds. In the outer solar system, this has been achieved with ALMA on Jupiter \citep{Cavalie2021,Benmahi2021}, Saturn \citep{Benmahi2022,Benmahi2025,Cavalie2026}, Titan \citep{Lellouch2019}, and Neptune \citep{Carrion-Gonzalez2023}. However, the nominal spectral resolution of 1.1\,MHz, combined with the high projection factor resulting from the $\sim$60$^\circ$N sub-Earth latitude, is too limiting to enable the unambiguous detection of any jet despite a careful analysis. 

\begin{figure}[!h]
    \centering
    \includegraphics[width=14cm]{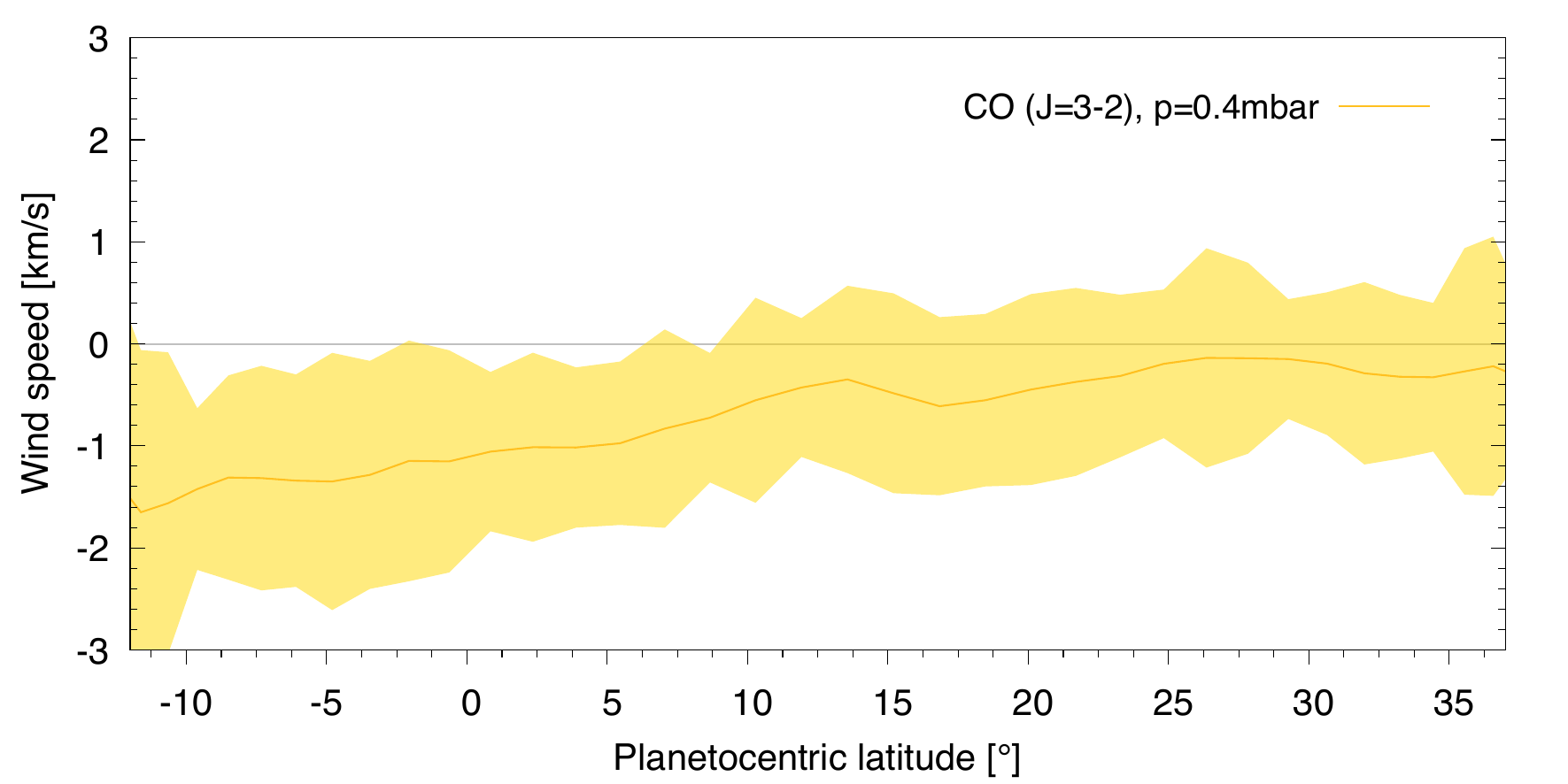}
    \caption{Wind speeds in the stratosphere of Uranus derived from the CO (J=3-2) observations. The yellow envelope represents the 1-$\sigma$ uncertainties. It shows the range of possible speeds at 0.4$^{+0.3}_{-0.1}$\,mbar (see \fig{fig:contribution_limb}) after averaging the two limbs and correcting for LOS projection effects caused by the large sub-Earth latitude and the relative longitude. Positive speeds correspond to prograde winds.}
    \label{fig:Uranus_winds}
  \end{figure}

From the CO observations in the combined-2022 dataset, we can only tentatively conclude that the envelope of possible zonal wind speeds, shown in \fig{fig:Uranus_winds}, seems incompatible with prograde winds between 10$^\circ$S and 10$^\circ$N at the probed pressure of 0.4$^{+0.3}_{-0.1}$\,mbar (see \fig{fig:contribution_limb}). This result is consistent with cloud-top measurements made in the upper troposphere \citep{Hammel2001,Hammel2005,Sromovsky2015,Sromovsky2024,Fletcher2020a}. Between 10$^\circ$N and 45$^\circ$N, the upper limits are 1000\ms~for retrograde winds and 500\ms~for prograde winds. 

The HCN data probes lower pressures than the CO ones, and could thus provide us with wind speed measurements at a different altitude level. However, using the HCN data from the combined-2022 cube does not provide any further constraint. Even though the HCN lines are narrower than the CO ones, enabling in principle more accurate Doppler wind measurements, the lower S/N in the HCN data precludes the derivation of better than an upper limit of $\pm$500\ms~over the probed latitude range.

Higher spectral resolution and sensitivity observations are needed to achieve wind speed measurements in the stratosphere of Uranus. Unfortunately, the situation regarding the sub-Earth latitude is not going to improve before the mid-2030s. This makes stratospheric wind measurements rather unlikely in the near future.
 
\section{Conclusion}\label{sec:Conclusion}
In this paper, we have presented spectral mapping observations of Uranus recorded in 2022 and 2024 with the ALMA interferometer. These observations targeted the CO (J=3-2) and HCN (J=4-3) rotational lines. The main results of this work can be summarized as follows:
\begin{itemize}
  \item A 5.8$\pm$0.3\,ppb tropospheric mole fraction of CO caused by an internal source is unambiguously detected for the first time as a broad absorption in the disk-center data. Applying a 1D thermochemical and diffusion model reproduces this abundance with a deep O/H ratio of 52$^{+30}_{-20}$, when assuming a constant vertical mixing in the range of 10$^7$-10$^9$\Kunit, with a nominal value of 10$^8$\Kunit. This range of O/H may be underestimated, because of other sources of uncertainties, especially the magnitude of vertical mixing at the high northern sub-observer latitude where the measurement was performed \citep{Wang2015}. These are investigated in a companion paper \citep{Briand2026}.

  \item When using the vertical profiles from \citet{Cavalie2014}, the model with an old impact of a comet provides us with the best match to the ALMA data compared to the IDP and local source models, because the comet profile produces broader lines than what the other models do, given the sluggish vertical mixing in the stratosphere of Uranus. The comet impact model produces the CO emission line in the 0.2-0.8\,mbar pressure range. A local source model scaled to match the ALMA data is also inconsistent with the earlier Herschel observations. We find a rather uniform horizontal distribution in the $\sim$15$^\circ$S-40$^\circ$N latitudinal range probed by the limb data, with an average column density of (7.5$\pm$0.6)\dix{15}\Coldens, when only counting the comet impact component of the vertical profile. Horizontal mixing timescales for an old comet impact are consistent with the impact date assumed when computing the vertical profile.
  
  \item A physical profile obtained with a vertical transport model and combining a 5.8\,ppb internal source with 2.4\dix{-7} of CO deposited by a comet at pressures lower than 0.1\,mbar 300\,years ago matches the whole dataset. The CO abundance peak was located at $\sim$0.8\,mbar at the time of the ALMA observations. CO thus has a dual origin in the atmosphere of Uranus, from its deep interior and from an old comet impact. 
  
  \item We detect HCN for the first time in the stratosphere of Uranus. The data can be reproduced with a constant mole fraction of (1.8$\pm$0.2)\dix{-11} restricted to pressures lower than 0.2\,mbar, placing the bulk of HCN above that of the stratospheric CO. Higher sensitivity and spectral resolution observations are needed to better constrain the HCN vertical profile and assess whether HCN has a source different from CO and if other processes, such as adsorption onto stratospheric aerosols, can explain the differences tentatively inferred in this work.
  
  \item Using a Doppler wind retrieval algorithm, we fail at detecting stratospheric winds from the CO and HCN data. With the limited S/N and spectral resolution of the data, we can only infer that the CO data probing the 0.4$^{+0.3}_{-0.1}$\,mbar level favors a retrograde flow between 10$^\circ$S and 10$^\circ$N.   
\end{itemize}

Unfortunately, the long orbital period of Uranus makes short-term prospects for improvements on various aspects of this study, such as the meridional distribution of stratospheric CO and zonal wind measurements, rather unlikely. With Uranus nearing its northern summer solstice, observations with ground-based and Earth-orbiting observatories will continue to be limited to the northern hemisphere in terms of latitudinal coverage for the next 10 years or so, and will be limited to equatorial latitudes for observations only probing along the planet limb. Only an orbital mission \citep{Arridge2014,Fletcher2020c,Cohen2022}, equipped with thermal infrared and/or submillimeter instrumentation, would ensure a full latitudinal coverage of the abundances of many stratospheric trace species and of zonal winds, whatever the season prevailing at the time of the mission. ALMA can nevertheless search for other species delivered by comet impacts, such as CS \citep{Marten1995,Moreno2017}. Also, continuum observation sampling the 0.1-1 mm wavelength range with ALMA may help to constrain the upper tropospheric H$_2$S abundance. However, an atmospheric entry probe \citep{Mousis2018,Mousis2022,Fletcher2022} is required to provide more context on the planet deep elemental and isotopic abundances. Mass spectrometry measurements of species (including disequilibrium species) present in the 10-20 bar pressure range complement the thermochemical models based on remote sensing observations of upper tropospheric composition \citep{Cavalie2020}. Such an ambitious mission still has to be selected and implemented by a space agency.

\section*{Acknowledgements}
T. Cavali\'e and C. Lefour were supported by the Programme National de Plan\'etologie (PNP) of CNRS/INSU and by the Centre National d'\'Etudes Spatiales (CNES). The authors acknowledge the support of the French Agence Nationale de la Recherche (ANR), under grant ANR-20-CE49-0009 (project SOUND). T. Cavali\'e thanks D. Petry from the ESO ALMA Regional Center for his assistance. Fletcher was supported by STFC Grant reference UKRI1205. 

This paper makes use of the following ALMA data: ADS/JAO.ALMA\#2021.1.01034.S and ADS/JAO.ALMA\#2022.1.00558.S. ALMA is a partnership of ESO (representing its member states), NSF (USA) and NINS (Japan), together with NRC (Canada), NSTC and ASIAA (Taiwan), and KASI (Republic of Korea), in cooperation with the Republic of Chile. The Joint ALMA Observatory is operated by ESO, AUI/NRAO and NAOJ.

\bibliographystyle{aa}

\begin{thebibliography}{117}
\expandafter\ifx\csname natexlab\endcsname\relax\def\natexlab#1{#1}\fi

\bibitem[{{Arridge} {et~al.}(2014){Arridge}, {Achilleos}, {Agarwal}, {Agnor},
  {Ambrosi}, {Andr{\'e}}, {Badman}, {Baines}, {Banfield}, {Barth{\'e}l{\'e}my},
  {Bisi}, {Blum}, {Bocanegra-Bahamon}, {Bonfond}, {Bracken}, {Brandt},
  {Briand}, {Briois}, {Brooks}, {Castillo-Rogez}, {Cavali{\'e}}, {Christophe},
  {Coates}, {Collinson}, {Cooper}, {Costa-Sitja}, {Courtin}, {Daglis}, {de
  Pater}, {Desai}, {Dirkx}, {Dougherty}, {Ebert}, {Filacchione}, {Fletcher},
  {Fortney}, {Gerth}, {Grassi}, {Grodent}, {Gr{\"u}n}, {Gustin}, {Hedman},
  {Helled}, {Henri}, {Hess}, {Hillier}, {Hofstadter}, {Holme}, {Horanyi},
  {Hospodarsky}, {Hsu}, {Irwin}, {Jackman}, {Karatekin}, {Kempf}, {Khalisi},
  {Konstantinidis}, {Kr{\"u}ger}, {Kurth}, {Labrianidis}, {Lainey}, {Lamy},
  {Laneuville}, {Lucchesi}, {Luntzer}, {MacArthur}, {Maier}, {Masters},
  {McKenna-Lawlor}, {Melin}, {Milillo}, {Moragas-Klostermeyer}, {Morschhauser},
  {Moses}, {Mousis}, {Nettelmann}, {Neubauer}, {Nordheim}, {Noyelles}, {Orton},
  {Owens}, {Peron}, {Plainaki}, {Postberg}, {Rambaux}, {Retherford}, {Reynaud},
  {Roussos}, {Russell}, {Rymer}, {Sallantin}, {S{\'a}nchez-Lavega}, {Santolik},
  {Saur}, {Sayanagi}, {Schenk}, {Schubert}, {Sergis}, {Sittler}, {Smith},
  {Spahn}, {Srama}, {Stallard}, {Sterken}, {Sternovsky}, {Tiscareno}, {Tobie},
  {Tosi}, {Trieloff}, {Turrini}, {Turtle}, {Vinatier}, {Wilson}, \&
  {Zarka}}]{Arridge2014}
{Arridge}, C.~S., {Achilleos}, N., {Agarwal}, J., {et~al.} 2014, \planss, 104,
  122

\bibitem[{{Atreya} {et~al.}(2020){Atreya}, {Hofstadter}, {In}, {Mousis}, {Reh},
  \& {Wong}}]{Atreya2020}
{Atreya}, S.~K., {Hofstadter}, M.~H., {In}, J.~H., {et~al.} 2020, \ssr, 216, 18

\bibitem[{{Baines} {et~al.}(1995){Baines}, {Mickelson}, {Larson}, \&
  {Ferguson}}]{Baines1995}
{Baines}, K.~H., {Mickelson}, M.~E., {Larson}, L.~E., \& {Ferguson}, D.~W.
  1995, \icarus, 114, 328

\bibitem[{{Benmahi} {et~al.}(2021){Benmahi}, {Cavali{\'e}}, {Greathouse},
  {Hue}, {Giles}, {Guerlet}, {Spiga}, \& {Cosentino}}]{Benmahi2021}
{Benmahi}, B., {Cavali{\'e}}, T., {Greathouse}, T.~K., {et~al.} 2021, \aap,
  652, A125

\bibitem[{{Benmahi} {et~al.}(2022){Benmahi}, {Cavali{\'e}}, {Fouchet},
  {Moreno}, {Lellouch}, {Bardet}, {Guerlet}, {Hue}, \& {Spiga}}]{Benmahi2022}
{Benmahi}, B., {Cavali{\'e}}, T., {Fouchet}, T., {et~al.} 2022, \aap, 666, A117

\bibitem[{{Benmahi} {et~al.}(2025){Benmahi}, {Cavali{\'e}}, {Fouchet},
  {Moreno}, {Lellouch}, {Bardet}, {Guerlet}, {Hue}, \& {Spiga}}]{Benmahi2025}
{Benmahi}, B., {Cavali{\'e}}, T., {Fouchet}, T., {et~al.} 2025, \aap, 696, C5

\bibitem[{{Bockel{\'e}e-Morvan} \& {Biver}(2017)}]{Bockelee2017}
{Bockel{\'e}e-Morvan}, D. \& {Biver}, N. 2017, Philosophical Transactions of
  the Royal Society of London Series A, 375, 20160252

\bibitem[{{Bolton} {et~al.}(2017){Bolton}, {Adriani}, {Adumitroaie}, {Allison},
  {Anderson}, {Atreya}, {Bloxham}, {Brown}, {Connerney}, {DeJong}, {Folkner},
  {Gautier}, {Grassi}, {Gulkis}, {Guillot}, {Hansen}, {Hubbard}, {Iess},
  {Ingersoll}, {Janssen}, {Jorgensen}, {Kaspi}, {Levin}, {Li}, {Lunine},
  {Miguel}, {Mura}, {Orton}, {Owen}, {Ravine}, {Smith}, {Steffes}, {Stone},
  {Stevenson}, {Thorne}, {Waite}, {Durante}, {Ebert}, {Greathouse}, {Hue},
  {Parisi}, {Szalay}, \& {Wilson}}]{Bolton2017}
{Bolton}, S.~J., {Adriani}, A., {Adumitroaie}, V., {et~al.} 2017, Science, 356,
  821

\bibitem[{{Borysow} {et~al.}(1985){Borysow}, {Trafton}, {Frommhold}, \&
  {Birnbaum}}]{Borysow1985}
{Borysow}, J., {Trafton}, L., {Frommhold}, L., \& {Birnbaum}, G. 1985, \apj,
  296, 644

\bibitem[{{Borysow} {et~al.}(1988){Borysow}, {Frommhold}, \&
  {Birnbaum}}]{Borysow1988}
{Borysow}, J., {Frommhold}, L., \& {Birnbaum}, G. 1988, \apj, 326, 509

\bibitem[{{Borysow} \& {Frommhold}(1986)}]{Borysow1986}
{Borysow}, A. \& {Frommhold}, L. 1986, \apj, 304, 849

\bibitem[{{Briand} {et~al.}(2026){Briand}, {Hue}, {Mousis}, {Cavali\'e},
  {Benest Couzinou}, {Schneeberger}, \& {Hofstadter}}]{Briand2026}
{Briand}, T., {Hue}, V., {Mousis}, O., {et~al.} 2026, \aap, submitted

\bibitem[{{Burgdorf} {et~al.}(2006){Burgdorf}, {Orton}, {van Cleve}, {Meadows},
  \& {Houck}}]{Burgdorf2006}
{Burgdorf}, M., {Orton}, G., {van Cleve}, J., {Meadows}, V., \& {Houck}, J.
  2006, \icarus, 184, 634

\bibitem[{{Carri{\'o}n-Gonz{\'a}lez} {et~al.}(2023){Carri{\'o}n-Gonz{\'a}lez},
  {Moreno}, {Lellouch}, {Cavali{\'e}}, {Guerlet}, {Milcareck}, {Spiga},
  {Cl{\'e}ment}, \& {Leconte}}]{Carrion-Gonzalez2023}
{Carri{\'o}n-Gonz{\'a}lez}, {\'O}., {Moreno}, R., {Lellouch}, E., {et~al.}
  2023, \aap, 674, L3

\bibitem[{{Cavali{\'e}} {et~al.}(2009){Cavali{\'e}}, {Billebaud}, {Dobrijevic},
  {Fouchet}, {Lellouch}, {Encrenaz}, {Brillet}, {Moriarty-Schieven},
  {Wouterloot}, \& {Hartogh}}]{Cavalie2009}
{Cavali{\'e}}, T., {Billebaud}, F., {Dobrijevic}, M., {et~al.} 2009, \icarus,
  203, 531

\bibitem[{{Cavali{\'e}} {et~al.}(2010){Cavali{\'e}}, {Hartogh}, {Billebaud},
  {Dobrijevic}, {Fouchet}, {Lellouch}, {Encrenaz}, {Brillet}, \&
  {Moriarty-Schieven}}]{Cavalie2010}
{Cavali{\'e}}, T., {Hartogh}, P., {Billebaud}, F., {et~al.} 2010, \aap, 510,
  A88

\bibitem[{{Cavali{\'e}} {et~al.}(2013){Cavali{\'e}}, {Feuchtgruber},
  {Lellouch}, {de Val-Borro}, {Jarchow}, {Moreno}, {Hartogh}, {Orton},
  {Greathouse}, {Billebaud}, {Dobrijevic}, {Lara}, {Gonz{\'a}lez}, \&
  {Sagawa}}]{Cavalie2013}
{Cavali{\'e}}, T., {Feuchtgruber}, H., {Lellouch}, E., {et~al.} 2013, \aap,
  553, A21

\bibitem[{{Cavali{\'e}} {et~al.}(2014){Cavali{\'e}}, {Moreno}, {Lellouch},
  {Hartogh}, {Venot}, {Orton}, {Jarchow}, {Encrenaz}, {Selsis}, {Hersant}, \&
  {Fletcher}}]{Cavalie2014}
{Cavali{\'e}}, T., {Moreno}, R., {Lellouch}, E., {et~al.} 2014, \aap, 562, A33

\bibitem[{{Cavali{\'e}} {et~al.}(2017){Cavali{\'e}}, {Venot}, {Selsis},
  {Hersant}, {Hartogh}, \& {Leconte}}]{Cavalie2017}
{Cavali{\'e}}, T., {Venot}, O., {Selsis}, F., {et~al.} 2017, \icarus, 291, 1

\bibitem[{{Cavali{\'e}} {et~al.}(2019){Cavali{\'e}}, {Hue}, {Hartogh},
  {Moreno}, {Lellouch}, {Feuchtgruber}, {Jarchow}, {Cassidy}, {Fletcher},
  {Billebaud}, {Dobrijevic}, {Rezac}, {Orton}, {Rengel}, {Fouchet}, \&
  {Guerlet}}]{Cavalie2019}
{Cavali{\'e}}, T., {Hue}, V., {Hartogh}, P., {et~al.} 2019, \aap, 630, A87

\bibitem[{{Cavali{\'e}} {et~al.}(2020){Cavali{\'e}}, {Venot}, {Miguel},
  {Fletcher}, {Wurz}, {Mousis}, {Bounaceur}, {Hue}, {Leconte}, \&
  {Dobrijevic}}]{Cavalie2020}
{Cavali{\'e}}, T., {Venot}, O., {Miguel}, Y., {et~al.} 2020, \ssr, 216, 58

\bibitem[{{Cavali{\'e}} {et~al.}(2021){Cavali{\'e}}, {Benmahi}, {Hue},
  {Moreno}, {Lellouch}, {Fouchet}, {Hartogh}, {Rezac}, {Greathouse},
  {Gladstone}, {Sinclair}, {Dobrijevic}, {Billebaud}, \&
  {Jarchow}}]{Cavalie2021}
{Cavali{\'e}}, T., {Benmahi}, B., {Hue}, V., {et~al.} 2021, \aap, 647, L8

\bibitem[{{Cavali{\'e}} {et~al.}(2023{\natexlab{a}}){Cavali{\'e}}, {Lunine}, \&
  {Mousis}}]{Cavalie2023a}
{Cavali{\'e}}, T., {Lunine}, J., \& {Mousis}, O. 2023{\natexlab{a}}, \natastron 7, 678

\bibitem[{{Cavali{\'e}} {et~al.}(2023{\natexlab{b}}){Cavali{\'e}}, {Rezac},
  {Moreno}, {Lellouch}, {Fouchet}, {Benmahi}, {Greathouse}, {Sinclair}, {Hue},
  {Hartogh}, {Dobrijevic}, {Carrasco}, \& {Perrin}}]{Cavalie2023b}
{Cavali{\'e}}, T., {Rezac}, L., {Moreno}, R., {et~al.} 2023{\natexlab{b}},
  \natastron, 7, 1048

\bibitem[{{Cavali{\'e}} {et~al.}(2024){Cavali{\'e}}, {Lunine}, {Mousis}, \&
  {Hueso}}]{Cavalie2024}
{Cavali{\'e}}, T., {Lunine}, J., {Mousis}, O., \& {Hueso}, R. 2024, \ssr, 220,
  8

\bibitem[{{Cavali{\'e}} {et~al.}(2026){Cavali{\'e}}, {Moreno}, {Lefour},
  {Benmahi}, {Fouchet}, {Lellouch}, {Ducreux}, {Gurwell}, {Gueth}, {Fletcher},
  \& {Bardet}}]{Cavalie2026}
{Cavali{\'e}}, T., {Moreno}, R., {Lefour}, C., {et~al.} 2026, \aap, 707, A240

\bibitem[{{Cl{\'e}ment} {et~al.}(2024){Cl{\'e}ment}, {Leconte}, {Spiga},
  {Guerlet}, {Selsis}, {Milcareck}, {Teinturier}, {Cavali{\'e}}, {Moreno}, {Lellouch},
  \& {Carr{\'i}on-Gonz{\'a}lez}}]{Clement2024}
{Cl{\'e}ment}, N., {Leconte}, J., {Spiga}, A., {et~al.} 2024, \aap, 690, A227

\bibitem[{{Cohen} {et~al.}(2022){Cohen}, {Beddingfield}, {Chancia},
  {DiBraccio}, {Hedman}, {MacKenzie}, {Mauk}, {Sayanagi}, {Soderlund},
  {Turtle}, {Ahrens}, {Arridge}, {Brooks}, {Bunce}, {Charnoz}, {Coustenis},
  {Dillman}, {Dutta}, {Fletcher}, {Harbison}, {Helled}, {Holme}, {Jozwiak},
  {Kasaba}, {Kollmann}, {Luszcz-Cook}, {Mandt}, {Mousis}, {Mura}, {Murakami},
  {Parisi}, {Rymer}, {Stanley}, {Stephan}, {Vervack}, {Wong}, \&
  {Wurz}}]{Cohen2022}
{Cohen}, I.~J., {Beddingfield}, C., {Chancia}, R., {et~al.} 2022, \psj, 3, 58

\bibitem[{{Conrath} {et~al.}(1987){Conrath}, {Hanel}, {Gautier}, {Marten}, \&
  {Lindal}}]{Conrath1987}
{Conrath}, B., {Hanel}, R., {Gautier}, D., {Marten}, A., \& {Lindal}, G. 1987,
  \jgr, 92, 15003

\bibitem[{{Coustenis} {et~al.}(1998){Coustenis}, {Salama}, {Lellouch},
  {Encrenaz}, {Bjoraker}, {Samuelson}, {de Graauw}, {Feuchtgruber}, \&
  {Kessler}}]{Coustenis1998}
{Coustenis}, A., {Salama}, A., {Lellouch}, E., {et~al.} 1998, \aap, 336, L85

\bibitem[{{de Kleer} {et~al.}(2015){de Kleer}, {Luszcz-Cook}, {de Pater},
  {{\'A}d{\'a}mkovics}, \& {Hammel}}]{deKleer2015}
{de Kleer}, K., {Luszcz-Cook}, S., {de Pater}, I., {{\'A}d{\'a}mkovics}, M., \&
  {Hammel}, H.~B. 2015, \icarus, 256, 120

\bibitem[{{de Pater} {et~al.}(2023){de Pater}, {Molter}, \&
  {Moeckel}}]{dePater2023b}
{de Pater}, I., {Molter}, E.~M., \& {Moeckel}, C.~M. 2023, Remote Sensing, 15,
  1313

\bibitem[{{de Pater} \& {Richmond}(1989)}]{dePater1989a}
{de Pater}, I. \& {Richmond}, M. 1989, \icarus, 80, 1

\bibitem[{{Dick} {et~al.}(2009){Dick}, {Drouin}, \& {Pearson}}]{Dick2009}
{Dick}, M.~J., {Drouin}, B.~J., \& {Pearson}, J.~C. 2009, \jqsrt, 110, 619

\bibitem[{{Encrenaz} {et~al.}(2004){Encrenaz}, {Lellouch}, {Drossart},
  {Feuchtgruber}, {Orton}, \& {Atreya}}]{Encrenaz2004}
{Encrenaz}, T., {Lellouch}, E., {Drossart}, P., {et~al.} 2004, \aap, 413, L5

\bibitem[{{Feuchtgruber} {et~al.}(1997){Feuchtgruber}, {Lellouch}, {de Graauw},
  {B{\'e}zard}, {Encrenaz}, \& {Griffin}}]{Feuchtgruber1997}
{Feuchtgruber}, H., {Lellouch}, E., {de Graauw}, T., {et~al.} 1997, \nat, 389,
  159

\bibitem[{{Flasar} {et~al.}(1987){Flasar}, {Conrath}, {Gierasch}, \&
  {Pirraglia}}]{Flasar1987}
{Flasar}, F.~M., {Conrath}, B.~J., {Gierasch}, P.~J., \& {Pirraglia}, J.~A.
  1987, \jgr, 92, 15011

\bibitem[{{Fletcher} {et~al.}(2020{\natexlab{a}}){Fletcher}, {de Pater},
  {Orton}, {Hofstadter}, {Irwin}, {Roman}, \& {Toledo}}]{Fletcher2020a}
{Fletcher}, L.~N., {de Pater}, I., {Orton}, G.~S., {et~al.} 2020{\natexlab{a}},
  \ssr, 216, 21

\bibitem[{{Fletcher} {et~al.}(2020{\natexlab{b}}){Fletcher}, {Helled},
  {Roussos}, {Jones}, {Charnoz}, {Andr{\'e}}, {Andrews}, {Bannister}, {Bunce},
  {Cavali{\'e}}, {Ferri}, {Fortney}, {Grassi}, {Griton}, {Hartogh}, {Hueso},
  {Kaspi}, {Lamy}, {Masters}, {Melin}, {Moses}, {Mousis}, {Nettleman},
  {Plainaki}, {Schmidt}, {Simon}, {Tobie}, {Tortora}, {Tosi}, \&
  {Turrini}}]{Fletcher2020c}
{Fletcher}, L.~N., {Helled}, R., {Roussos}, E., {et~al.} 2020{\natexlab{b}},
  \planss, 191, 105030

\bibitem[{{Fletcher} {et~al.}(2020{\natexlab{c}}){Fletcher}, {Kaspi},
  {Guillot}, \& {Showman}}]{Fletcher2020b}
{Fletcher}, L.~N., {Kaspi}, Y., {Guillot}, T., \& {Showman}, A.~P.
  2020{\natexlab{c}}, \ssr, 216, 30

\bibitem[{{Fletcher} {et~al.}(2022){Fletcher}, {Helled}, {Roussos}, {Jones},
  {Charnoz}, {Andr{\'e}}, {Andrews}, {Bannister}, {Bunce}, {Cavali{\'e}},
  {Ferri}, {Fortney}, {Grassi}, {Griton}, {Hartogh}, {Hueso}, {Kaspi}, {Lamy},
  {Masters}, {Melin}, {Moses}, {Mousis}, {Nettleman}, {Plainaki}, {Schmidt},
  {Simon}, {Tobie}, {Tortora}, {Tosi}, \& {Turrini}}]{Fletcher2022}
{Fletcher}, L.~N., {Helled}, R., {Roussos}, E., {et~al.} 2022, Experimental
  Astronomy, 54, 1015

\bibitem[{{Fray} \& {Schmitt}(2009)}]{Fray2009}
{Fray}, N. \& {Schmitt}, B. 2009, \planss, 57, 2053

\bibitem[{{Guillot}(2005)}]{Guillot2005}
{Guillot}, T. 2005, \areps, 33, 493

\bibitem[{{Guillot} {et~al.}(2023){Guillot}, {Fletcher}, {Helled}, {Ikoma},
  {Line}, \& {Paramentier}}]{Guillot2023}
{Guillot}, T., {Fletcher}, L.~N., {Helled}, R., {et~al.} 2023, in Astronomical
  Society of the Pacific Conference Series, Vol. 534, Protostars and Planets
  VII, ed. S.~{Inutsuka}, Y.~{Aikawa}, T.~{Muto}, K.~{Tomida}, \& M.~{Tamura},
  947

\bibitem[{{Hammel} {et~al.}(2001){Hammel}, {Rages}, {Lockwood}, {Karkoschka},
  \& {de Pater}}]{Hammel2001}
{Hammel}, H.~B., {Rages}, K., {Lockwood}, G.~W., {Karkoschka}, E., \& {de
  Pater}, I. 2001, \icarus, 153, 229

\bibitem[{{Hammel} {et~al.}(2005){Hammel}, {de Pater}, {Gibbard}, {Lockwood},
  \& {Rages}}]{Hammel2005}
{Hammel}, H.~B., {de Pater}, I., {Gibbard}, S., {Lockwood}, G.~W., \& {Rages},
  K. 2005, \icarus, 175, 534

\bibitem[{{Hartogh} {et~al.}(2011){Hartogh}, {Lellouch}, {Moreno},
  {Bockel{\'e}e-Morvan}, {Biver}, {Cassidy}, {Rengel}, {Jarchow},
  {Cavali{\'e}}, {Crovisier}, {Helmich}, \& {Kidger}}]{Hartogh2011a}
{Hartogh}, P., {Lellouch}, E., {Moreno}, R., {et~al.} 2011, \aap, 532, L2

\bibitem[{{Helled} \& {Fortney}(2020)}]{Helled2020b}
{Helled}, R. \& {Fortney}, J.~J. 2020, Philosophical Transactions of the Royal
  Society of London Series A, 378, 20190474

\bibitem[{{Helled} {et~al.}(2020){Helled}, {Nettelmann}, \&
  {Guillot}}]{Helled2020a}
{Helled}, R., {Nettelmann}, N., \& {Guillot}, T. 2020, \ssr, 216, 38

\bibitem[{{Hesman} {et~al.}(2007){Hesman}, {Davis}, {Matthews}, \&
  {Orton}}]{Hesman2007}
{Hesman}, B.~E., {Davis}, G.~R., {Matthews}, H.~E., \& {Orton}, G.~S. 2007,
  \icarus, 186, 342

\bibitem[{{Hue} {et~al.}(2018){Hue}, {Hersant}, {Cavali{\'e}}, {Dobrijevic}, \&
  {Sinclair}}]{Hue2018}
{Hue}, V., {Hersant}, F., {Cavali{\'e}}, T., {Dobrijevic}, M., \& {Sinclair},
  J.~A. 2018, \icarus, 307, 106

\bibitem[{{Hueso} {et~al.}(2020){Hueso}, {Guillot}, \&
  {S{\'a}nchez-Lavega}}]{Hueso2020}
{Hueso}, R., {Guillot}, T., \& {S{\'a}nchez-Lavega}, A. 2020, Philosophical
  Transactions of the Royal Society of London Series A, 378, 20190476

\bibitem[{{Hyder} {et~al.}(2025){Hyder}, {Li}, {Chanover}, \&
  {Bjoraker}}]{Hyder2025}
{Hyder}, A., {Li}, C., {Chanover}, N., \& {Bjoraker}, G. 2025, Nature
  Astronomy, 9, 211

\bibitem[{{Irwin} {et~al.}(2018){Irwin}, {Toledo}, {Garland}, {Teanby},
  {Fletcher}, {Orton}, \& {B{\'e}zard}}]{Irwin2018}
{Irwin}, P. G.~J., {Toledo}, D., {Garland}, R., {et~al.} 2018, Nature
  Astronomy, 2, 420

\bibitem[{{Irwin} {et~al.}(2019){Irwin}, {Toledo}, {Braude}, {Bacon},
  {Weilbacher}, {Teanby}, {Fletcher}, \& {Orton}}]{Irwin2019b}
{Irwin}, P. G.~J., {Toledo}, D., {Braude}, A.~S., {et~al.} 2019, \icarus, 331,
  69

\bibitem[{{Irwin} {et~al.}(2025){Irwin}, {Wenkert}, {Simon}, {Dahl}, \&
  {Hammel}}]{Irwin2025}
{Irwin}, P. G.~J., {Wenkert}, D.~D., {Simon}, A.~A., {Dahl}, E., \& {Hammel},
  H.~B. 2025, \mnras, 540, 1719

\bibitem[{{Karkoschka} \& {Tomasko}(2009)}]{Karkoschka2009}
{Karkoschka}, E. \& {Tomasko}, M. 2009, \icarus, 202, 287

\bibitem[{{Karkoschka} \& {Tomasko}(2011)}]{Karkoschka2011}
{Karkoschka}, E. \& {Tomasko}, M.~G. 2011, \icarus, 211, 780

\bibitem[{{Landgraf} {et~al.}(2002){Landgraf}, {Liou}, {Zook}, \&
  {Gr{\"u}n}}]{Landgraf2002}
{Landgraf}, M., {Liou}, J.-C., {Zook}, H.~A., \& {Gr{\"u}n}, E. 2002, \aj, 123,
  2857

\bibitem[{{Leconte} {et~al.}(2017){Leconte}, {Selsis}, {Hersant}, \&
  {Guillot}}]{Leconte2017}
{Leconte}, J., {Selsis}, F., {Hersant}, F., \& {Guillot}, T. 2017, \aap, 598,
  A98

\bibitem[{{Lefour} {et~al.}(2026){Lefour}, {Cavali{\'e}}, {Moreno}, {Rezac},
  {Fouchet}, {Lellouch}, \& {Hartogh}}]{Lefour2026}
{Lefour}, C., {Cavali{\'e}}, T., {Moreno}, R., {et~al.} 2026, \aap, DOI: 10.1051/0004-6361/202659264

\bibitem[{{Lellouch} {et~al.}(1995){Lellouch}, {Paubert}, {Moreno}, {Festou},
  {Bezard}, {Bockelee-Morvan}, {Colom}, {Crovisier}, {Encrenaz}, {Gautier},
  {Marten}, {Despois}, {Strobel}, \& {Sievers}}]{Lellouch1995}
{Lellouch}, E., {Paubert}, G., {Moreno}, R., {et~al.} 1995, \nat, 373, 592

\bibitem[{{Lellouch} {et~al.}(2002){Lellouch}, {B{\'e}zard}, {Moses}, {Davis},
  {Drossart}, {Feuchtgruber}, {Bergin}, {Moreno}, \& {Encrenaz}}]{Lellouch2002}
{Lellouch}, E., {B{\'e}zard}, B., {Moses}, J.~I., {et~al.} 2002, \icarus, 159,
  112

\bibitem[{{Lellouch} {et~al.}(2005){Lellouch}, {Moreno}, \&
  {Paubert}}]{Lellouch2005}
{Lellouch}, E., {Moreno}, R., \& {Paubert}, G. 2005, \aap, 430, L37

\bibitem[{{Lellouch} {et~al.}(2006){Lellouch}, {B{\'e}zard}, {Strobel},
  {Bjoraker}, {Flasar}, \& {Romani}}]{Lellouch2006}
{Lellouch}, E., {B{\'e}zard}, B., {Strobel}, D.~F., {et~al.} 2006, \icarus,
  184, 478

\bibitem[{{Lellouch} {et~al.}(2015){Lellouch}, {Moreno}, {Orton},
  {Feuchtgruber}, {Cavali{\'e}}, {Moses}, {Hartogh}, {Jarchow}, \&
  {Sagawa}}]{Lellouch2015a}
{Lellouch}, E., {Moreno}, R., {Orton}, G.~S., {et~al.} 2015, \aap, 579, A121

\bibitem[{{Lellouch} {et~al.}(2019){Lellouch}, {Gurwell}, {Moreno},
  {Vinatier}, {Strobel}, {Moullet}, {Butler}, {Lara}, {Hidayat}, \&
  {Villard}}]{Lellouch2019}
{Lellouch}, E., {Gurwell}, M., {Moreno}, R., {et~al.} 2019, \natastron, 3, 614

\bibitem[{{Li} {et~al.}(2020){Li}, {Ingersoll}, {Bolton}, {Levin}, {Janssen},
  {Atreya}, {Lunine}, {Steffes}, {Brown}, {Guillot}, {Allison}, {Arballo},
  {Bellotti}, {Adumitroaie}, {Gulkis}, {Hodges}, {Li}, {Misra}, {Orton},
  {Oyafuso}, {Santos-Costa}, {Waite}, \& {Zhang}}]{Li2020}
{Li}, C., {Ingersoll}, A., {Bolton}, S., {et~al.} 2020, \natastron, 4,
  609

\bibitem[{{Lindal} {et~al.}(1987){Lindal}, {Lyons}, {Sweetnam}, {Eshleman}, \&
  {Hinson}}]{Lindal1987}
{Lindal}, G.~F., {Lyons}, J.~R., {Sweetnam}, D.~N., {Eshleman}, V.~R., \&
  {Hinson}, D.~P. 1987, \jgr, 92, 14987

\bibitem[{{Lodders}(2021)}]{Lodders2021}
{Lodders}, K. 2021, \ssr, 217, 44

\bibitem[{{Luszcz-Cook} \& {de Pater}(2013)}]{Luszcz-Cook2013}
{Luszcz-Cook}, S.~H. \& {de Pater}, I. 2013, \icarus, 222, 379

\bibitem[{{Marten} {et~al.}(1995){Marten}, {Gautier}, {Griffin}, {Matthews},
  {Naylor}, {Davis}, {Owen}, {Orton}, {Bockel{\'e}e-Morvan}, {Colom},
  {Crovisier}, {Lellouch}, {de Pater}, {Atreya}, {Strobel}, {Han}, \&
  {Sanders}}]{Marten1995}
{Marten}, A., {Gautier}, D., {Griffin}, M.~J., {et~al.} 1995, \grl, 22, 1589

\bibitem[{{Milcareck} {et~al.}(2024){Milcareck}, {Guerlet}, {Montmessin},
  {Spiga}, {Leconte}, {Millour}, {Cl{\'e}ment}, {Fletcher}, {Roman},
  {Lellouch}, {Moreno}, {Cavali{\'e}}, \&
  {Carri{\'o}n-Gonz{\'a}lez}}]{Milcareck2024}
{Milcareck}, G., {Guerlet}, S., {Montmessin}, F., {et~al.} 2024, \aap, 686,
  A303

\bibitem[{{Molter} {et~al.}(2021){Molter}, {de Pater}, {Luszcz-Cook},
  {Tollefson}, {Sault}, {Butler}, \& {de Boer}}]{Molter2021}
{Molter}, E.~M., {de Pater}, I., {Luszcz-Cook}, S., {et~al.} 2021, \psj, 2, 3

\bibitem[{{Moreno} {et~al.}(2003){Moreno}, {Marten}, {Matthews}, \&
  {Biraud}}]{Moreno2003}
{Moreno}, R., {Marten}, A., {Matthews}, H.~E., \& {Biraud}, Y. 2003, \planss,
  51, 591

\bibitem[{{Moreno} {et~al.}(2005){Moreno}, {Marten}, \& {Hidayat}}]{Moreno2005}
{Moreno}, R., {Marten}, A., \& {Hidayat}, T. 2005, \aap, 437, 319

\bibitem[{{Moreno} {et~al.}(2017){Moreno}, {Lellouch}, {Cavali{\'e}}, \&
  {Moullet}}]{Moreno2017}
{Moreno}, R., {Lellouch}, E., {Cavali{\'e}}, T., \& {Moullet}, A. 2017, \aap,
  608, L5

\bibitem[{{Moses}(2014)}]{Moses2014}
{Moses}, J.~I. 2014, Philosophical Transactions of the Royal Society of London
  Series A, 372, 20130073

\bibitem[{{Moses} \& {Poppe}(2017)}]{Moses2017}
{Moses}, J.~I. \& {Poppe}, A.~R. 2017, \icarus, 297, 33

\bibitem[{{Moses} {et~al.}(2000){Moses}, {Lellouch}, {B{\'e}zard}, {Gladstone},
  {Feuchtgruber}, \& {Allen}}]{Moses2000b}
{Moses}, J.~I., {Lellouch}, E., {B{\'e}zard}, B., {et~al.} 2000, \icarus, 145,
  166

\bibitem[{{Moses} {et~al.}(2018){Moses}, {Fletcher}, {Greathouse}, {Orton}, \&
  {Hue}}]{Moses2018}
{Moses}, J.~I., {Fletcher}, L.~N., {Greathouse}, T.~K., {Orton}, G.~S., \&
  {Hue}, V. 2018, \icarus, 307, 124

\bibitem[{{Moses} {et~al.}(2023){Moses}, {Brown}, {Koskinen}, {Fletcher}, 
  {Serigano}, {Guerlet}, {Moore}, {Waite}, {Ben-Jaffel}, {Galand}, {Chadney}, 
  {H{\"o}rst}, {Sinclair}, {Vuitton}, \& {M{\"u}ller-Wodarg}}]{Moses2023}
{Moses}, J.~I., {Brown}, Z.~L., {Koskinen}, T.~T., {et~al.} 2023, \icarus, 391,
  115328

\bibitem[{{Mousis} {et~al.}(2018){Mousis}, {Atkinson}, {Cavali{\'e}},
  {Fletcher}, {Amato}, {Aslam}, {Ferri}, {Renard}, {Spilker}, {Venkatapathy},
  {Wurz}, {Aplin}, {Coustenis}, {Deleuil}, {Dobrijevic}, {Fouchet}, {Guillot},
  {Hartogh}, {Hewagama}, {Hofstadter}, {Hue}, {Hueso}, {Lebreton}, {Lellouch},
  {Moses}, {Orton}, {Pearl}, {S{\'a}nchez-Lavega}, {Simon}, {Venot}, {Waite},
  {Achterberg}, {Atreya}, {Billebaud}, {Blanc}, {Borget}, {Brugger}, {Charnoz},
  {Chiavassa}, {Cottini}, {d'Hendecourt}, {Danger}, {Encrenaz}, {Gorius},
  {Jorda}, {Marty}, {Moreno}, {Morse}, {Nixon}, {Reh}, {Ronnet}, {Schmider},
  {Sheridan}, {Sotin}, {Vernazza}, \& {Villanueva}}]{Mousis2018}
{Mousis}, O., {Atkinson}, D.~H., {Cavali{\'e}}, T., {et~al.} 2018, \planss,
  155, 12

\bibitem[{{Mousis} {et~al.}(2020){Mousis}, {Aguichine}, {Helled}, {Irwin}, \&
  {Lunine}}]{Mousis2020}
{Mousis}, O., {Aguichine}, A., {Helled}, R., {Irwin}, P.~G.~J., \& {Lunine},
  J.~I. 2020, Philosophical Transactions of the Royal Society of London Series
  A, 378, 20200107

\bibitem[{{Mousis} {et~al.}(2022){Mousis}, {Atkinson}, {Ambrosi}, {Atreya},
  {Banfield}, {Barabash}, {Blanc}, {Cavali{\'e}}, {Coustenis}, {Deleuil},
  {Durry}, {Ferri}, {Fletcher}, {Fouchet}, {Guillot}, {Hartogh}, {Hueso},
  {Hofstadter}, {Lebreton}, {Mandt}, {Rauer}, {Rannou}, {Renard},
  {S{\'a}nchez-Lavega}, {Sayanagi}, {Simon}, {Spilker}, {Venkatapathy},
  {Waite}, \& {Wurz}}]{Mousis2022}
{Mousis}, O., {Atkinson}, D.~H., {Ambrosi}, R., {et~al.} 2022, Experimental
  Astronomy, 54, 975

\bibitem[{{Orton} {et~al.}(2014{\natexlab{a}}){Orton}, {Fletcher}, {Moses},
  {Mainzer}, {Hines}, {Hammel}, {Martin-Torres}, {Burgdorf}, {Merlet}, \&
  {Line}}]{Orton2014a}
{Orton}, G.~S., {Fletcher}, L.~N., {Moses}, J.~I., {et~al.} 2014{\natexlab{a}},
  \icarus, 243, 494

\bibitem[{{Orton} {et~al.}(2014{\natexlab{b}}){Orton}, {Moses}, {Fletcher},
  {Mainzer}, {Hines}, {Hammel}, {Martin-Torres}, {Burgdorf}, {Merlet}, \&
  {Line}}]{Orton2014b}
{Orton}, G.~S., {Moses}, J.~I., {Fletcher}, L.~N., {et~al.} 2014{\natexlab{b}},
  \icarus, 243, 471

\bibitem[{Perry {et~al.}(2018)Perry, Waite~Jr., Mitchell, Miller, Cravens,
  Perryman, Moore, Yelle, Hsu, Hedman, Cuzzi, Strobel, Hamil, Glein, Paxton,
  Teolis, \& McNutt~Jr.}]{Perry2018}
Perry, M.~E., Waite~Jr., J.~H., Mitchell, D.~G., {et~al.} 2018, \grl, 45,
  10,093

\bibitem[{{Pety}(2005)}]{Pety2005}
{Pety}, J. 2005, in SF2A-2005: Semaine de l'Astrophysique Francaise, ed.
  F.~{Casoli}, T.~{Contini}, J.~M. {Hameury}, \& L.~{Pagani}, 721

\bibitem[{{Pickett} {et~al.}(1998){Pickett}, {Poynter}, {Cohen}, {Delitsky},
  {Pearson}, \& {M{\"u}ller}}]{Pickett1998}
{Pickett}, H.~M., {Poynter}, R.~L., {Cohen}, E.~A., {et~al.} 1998, \jqsrt, 60,
  883

\bibitem[{{Poppe}(2016)}]{Poppe2016}
{Poppe}, A.~R. 2016, \icarus, 264, 369

\bibitem[{{Prang{\'e}} {et~al.}(2006){Prang{\'e}}, {Fouchet}, {Courtin},
  {Connerney}, \& {McConnell}}]{Prange2006}
{Prang{\'e}}, R., {Fouchet}, T., {Courtin}, R., {Connerney}, J.~E.~P., \&
  {McConnell}, J.~C. 2006, \icarus, 180, 379

\bibitem[{{Reinhardt} {et~al.}(2020){Reinhardt}, {Chau}, {Stadel}, \&
  {Helled}}]{Reinhardt2020}
{Reinhardt}, C., {Chau}, A., {Stadel}, J., \& {Helled}, R. 2020, \mnras, 492,
  5336

\bibitem[{Rohart {et~al.}(1987)Rohart, Derozier, \& Legrand}]{Rohart1987}
Rohart, F., Derozier, D., \& Legrand, J. 1987, \jchemphys, 87, 5794

\bibitem[{{Roman} {et~al.}(2018){Roman}, {Banfield}, \& {Gierasch}}]{Roman2018}
{Roman}, M.~T., {Banfield}, D., \& {Gierasch}, P.~J. 2018, \icarus, 310, 54

\bibitem[{{Roman} {et~al.}(2026){Roman}, {Fletcher}, {Hammel}, {King}, {Irwin},
  {de Pater}, {Moses}, {Orton}, {Rowe-Gurney}, {Melin}, {Hedman}, {Toogood},
  {Harkett}, {Penn}, \& {Milam}}]{Roman2026}
{Roman}, M.~T., {Fletcher}, L.~N., {Hammel}, H.~B., {et~al.} 2026, \natcom, submitted

\bibitem[{{Solem} (1994){Solem}}]{Solem1994}
{Solem}, J.~C. 1994, \nat, 370, 349

\bibitem[{{Sromovsky} {et~al.}(2011){Sromovsky}, {Fry}, \&
  {Kim}}]{Sromovsky2011}
{Sromovsky}, L.~A., {Fry}, P.~M., \& {Kim}, J.~H. 2011, \icarus, 215, 292

\bibitem[{{Sromovsky} {et~al.}(2014){Sromovsky}, {Karkoschka}, {Fry}, {Hammel},
  {de Pater}, \& {Rages}}]{Sromovsky2014}
{Sromovsky}, L.~A., {Karkoschka}, E., {Fry}, P.~M., {et~al.} 2014, \icarus,
  238, 137

\bibitem[{{Sromovsky} {et~al.}(2015){Sromovsky}, {de Pater}, {Fry}, {Hammel},
  \& {Marcus}}]{Sromovsky2015}
{Sromovsky}, L.~A., {de Pater}, I., {Fry}, P.~M., {Hammel}, H.~B., \& {Marcus},
  P. 2015, \icarus, 258, 192

\bibitem[{{Sromovsky} {et~al.}(2024){Sromovsky}, {Fry}, {de Pater}, \&
  {Hammel}}]{Sromovsky2024}
{Sromovsky}, L.~A., {Fry}, P.~M., {de Pater}, I., \& {Hammel}, H.~B. 2024,
  \icarus, 420, 116186

\bibitem[{{Sromovsky} \& {Fry}(2008)}]{Sromovsky2008}
{Sromovsky}, L.~A. \& {Fry}, P.~M. 2008, \icarus, 193, 252

\bibitem[{{Strobel} \& {Yung}(1979)}]{Strobel1979}
{Strobel}, D.~F. \& {Yung}, Y.~L. 1979, \icarus, 37, 256

\bibitem[{Teanby \& Irwin(2013)}]{Teanby2013}
Teanby, N.~A. \& Irwin, P. G.~J. 2013, \apjl, 775, L49

\bibitem[{{Teanby} {et~al.}(2019){Teanby}, {Irwin}, \&{Moses}}]{Teanby2019}
{Teanby}, N.~A., {Irwin}, P.~G.~J., \& {Moses}, J.~I. 2019,
  \icarus, 319, 86

\bibitem[{{Teanby} {et~al.}(2020){Teanby}, {Irwin}, {Moses}, \&
  {Helled}}]{Teanby2020}
{Teanby}, N.~A., {Irwin}, P.~G.~J., {Moses}, J.~I., \& {Helled}, R. 2020,
  \ptrslsA, 378, 20190489

\bibitem[{{Teanby} {et~al.}(2022){Teanby}, {Irwin}, {Sylvestre}, {Nixon}, \&
  {Cordiner}}]{Teanby2022}
{Teanby}, N.~A., {Irwin}, P.~G.~J., {Sylvestre}, M., {Nixon}, C.~A., \&
  {Cordiner}, M.~A. 2022, \psj, 3, 96

\bibitem[{{Tollefson} {et~al.}(2021){Tollefson}, {de Pater}, {Molter}, {Sault},
  {Butler}, {Luszcz-Cook}, \& {DeBoer}}]{Tollefson2021}
{Tollefson}, J., {de Pater}, I., {Molter}, E.~M., {et~al.} 2021, \psj, 2, 105

\bibitem[{{Vazan} \& {Helled}(2020)}]{Vazan2020}
{Vazan}, A. \& {Helled}, R. 2020, \aap, 633, A50

\bibitem[{{Venot} {et~al.}(2019){Venot}, {Bounaceur}, {Dobrijevic},
  {H{\'e}brard}, {Cavali{\'e}}, {Tremblin}, {Drummond}, \&
  {Charnay}}]{Venot2019}
{Venot}, O., {Bounaceur}, R., {Dobrijevic}, M., {et~al.} 2019, \aap, 624, A58

\bibitem[{{Venot} {et~al.}(2020){Venot}, {Cavali{\'e}}, {Bounaceur},
  {Tremblin}, {Brouillard}, \& {Lhoussaine Ben Brahim}}]{Venot2020}
{Venot}, O., {Cavali{\'e}}, T., {Bounaceur}, R., {et~al.} 2020, \aap, 634, A78

\bibitem[{{Visscher} {et~al.}(2010){Visscher}, {Moses}, \&
  {Saslow}}]{Visscher2010}
{Visscher}, C., {Moses}, J.~I., \& {Saslow}, S.~A. 2010, \icarus, 209, 602

\bibitem[{{Waite} {et~al.}(2018){Waite}, {Perryman}, {Perry}, {Miller}, {Bell},
  {Glein}, {Grimes}, {Hedman}, {Cuzzi}, {Brockwell}, {Teolis}, {Moore},
  {Mitchell}, {Persoon}, {Kurth}, {Wahlund}, {Morooka}, {Hadid}, {Walker},
  {Nagy}, {Yelle}, {Ledvina}, {Johnson}, {Tseng}, {Tucker}, \&
  {Ip}}]{Waite2018}
{Waite}, J.~H., {Perryman}, R.~S., {Perry}, M.~E., {et~al.} 2018, \science,
  362, 51

\bibitem[{{Wang} {et~al.}(2015){Wang}, {Gierasch}, {Lunine}, \&
  {Mousis}}]{Wang2015}
{Wang}, D., {Gierasch}, P.~J., {Lunine}, J.~I., \& {Mousis}, O. 2015, \icarus,
  250, 154

\bibitem[{{Wang} {et~al.}(2025){Wang}, {Li}, {Roman}, {Zhang}, {Jiang}, {Fry},
  {Li}, {Milcareck}, {Sanchez-Lavega}, {Perez-Hoyos}, {Hueso}, {Guillot},
  {Nixon}, {Dyudina}, {West}, \& {Kenyon}}]{Wang2025}
{Wang}, X., {Li}, L., {Roman}, M., {et~al.} 2025, \grl, 52, e2025GL115660

\bibitem[{{Zahnle} {et~al.}(2003){Zahnle}, {Schenk}, {Levison}, \&
  {Dones}}]{Zahnle2003}
{Zahnle}, K., {Schenk}, P., {Levison}, H., \& {Dones}, L. 2003, \icarus, 163,
  263

\bibitem[{{Zhu} \& {Dong}(2021)}]{Zhu2021}
{Zhu}, W. \& {Dong}, S. 2021, \araa, 59, 291

\end{thebibliography}

\end{document}